\documentclass{aa}  
\usepackage{graphicx}
\usepackage{txfonts}
\usepackage{natbib}
\usepackage[colorlinks=true,linkcolor=blue,citecolor=blue,urlcolor=blue]{hyperref}
\usepackage{caption}
\usepackage{booktabs}  
\usepackage{siunitx} 
\usepackage{gensymb}
\usepackage{amsmath} 
\usepackage{breqn}
\usepackage{amssymb}
\usepackage{graphicx}
\usepackage{float} 

\bibpunct{(}{)}{;}{a}{}{,} 

\defcitealias{2022florian}{FN22}
\defcitealias{2022thomas}{TS22}
\defcitealias{2021GaiaMC}{G21}

\begin{document} 

   \title{The VMC survey – LIV. The internal kinematics of the Large Magellanic Cloud with new VISTA observations}

   \titlerunning{LMC internal kinematics}

   \author{Sreepriya Vijayasree,\inst{1,2}
   \fnmsep\thanks{E-mail: \href{mailto:svijayasree@aip.de}{svijayasree@aip.de}}
   Florian Niederhofer,\inst{2}
   Maria-Rosa L. Cioni,\inst{1,2}
   Lara Cullinane,\inst{2}
   Kenji Bekki,\inst{3}
   Jacco Th. van Loon,\inst{4}
   Nikolay Kacharov,\inst{2}
   Richard de Grijs,\inst{5,6,7}
   Valentin D. Ivanov,\inst{8}
   Joana M. Oliveira,\inst{4}
   Francesca Dresbach,\inst{4}
   Martin A. T. Groenewegen,\inst{9}
   and Denis Erkal\inst{10}
    }
    \authorrunning{Vijayasree et al.}
    
   \institute{Institut für Physik und Astronomie, Universität Potsdam, Haus 28, Karl-Liebknecht-Str. 24/25, D-14476 Golm (Potsdam), Germany
   \and
   Leibniz-Institut für Astrophysik Potsdam, An der Sternwarte 16, D-14482 Potsdam, Germany
   \and
   International Centre for Radio Astronomy Research, The University of Western Australia, 35 Stirling Highway, Crawley, Western Australia 6009, Australia
   \and
   Lennard-Jones Laboratories, Keele University, Keele ST5 5BG, UK
   \and
   School of Mathematical and Physical Sciences, Macquarie University, Balaclava Road, Sydney, NSW 2109, Australia
   \and
   Astrophysics and Space Technologies Research Centre, Macquarie University, Balaclava Road, Sydney, NSW 2109, Australia
   \and
   International Space Science Institute--Beijing, 1 Nanertiao, Zhongguancun, Hai Dian District, Beijing 100190, China
   \and
   European Southern Observatory, Karl-Schwarzschild-Str. 2, D-85748 Garching bei München, Germany
   \and
   Koninklijke Sterrenwacht van België, Ringlaan 3, B–1180 Brussels, Belgium
   \and
   Department of Physics, University of Surrey, Guildford GU2 7XH, UK
   }


  \abstract
   {The study of the internal kinematics of galaxies provides insights into their past evolution, current dynamics, and future trajectory. The Large Magellanic Cloud (LMC), as the largest and one of the nearest satellite galaxies of the Milky Way (MW), presents unique opportunities to investigate these phenomena in great detail.}
   {We aim to investigate the internal kinematics of the LMC by deriving precise stellar proper motions using data from the VISTA survey of the Magellanic Clouds system (VMC). The main objective is to refine the LMC’s dynamical parameters using improved proper motion measurements exploiting the additional epochs of observations from the VMC survey.
   }
   {We utilised high-precision proper motion measurements from the VMC survey, leveraging an extended time baseline from approximately 2 to 10 years. This extension significantly enhanced the precision of the proper motion data, reducing uncertainties from 6 mas yr$^{-1}$ in prior studies using the VMC dataset to 1.5 mas yr$^{-1}$. Using this data, we derived geometrical and kinematic parameters, and generated velocity maps and rotation curves in the LMC disc plane and the sky plane, for both young and old stellar populations. Finally, we compared a suite of dynamical models that simulate the interaction of the LMC with the MW and Small Magellanic Cloud (SMC), against the observations.}
   {The tangential rotation curve reveals an asymmetric drift between young and old stars, while the radial velocity curve for the young population shows an increasing trend within the inner bar region, suggesting non-circular orbits. The internal rotation map confirms the clockwise rotation around the dynamical centre of the LMC, which is consistent with previous predictions. A significant residual motion was detected towards the north-east of the LMC, directed away from the centre. This feature observed in the inner disc region is kinematically connected with a substructure identified in the periphery known as Eastern Substructure 1. This motion of the LMC sources suggests a possible tidal influence from the MW, combined with the effects of the recent close pericentre passage of the SMC $\sim$150 Myr ago.}
  {}
  

   \maketitle
%

\section{Introduction}

The Large Magellanic Cloud (LMC) is a barred spiral galaxy with irregular features, classified as SB(s)m in the galaxy classification scheme \citep{1972vaucouleurs}. It is part of a large system comprising an ensemble of structures including the LMC, the Small Magellanic Cloud (SMC), the Magellanic Bridge connecting the two galaxies \citep{1963hindman}, a vast gaseous stream known as the Magellanic Stream spanning nearly 200 deg across the southern sky \citep{1974mathewson, 2016donghia}, and the Leading Arm, which forms part of this gaseous stream located above the two galaxies \citep{2008nidever}. Approximately 50 kpc from the Milky Way (MW) \citep{2014degrijs, 2019pietrzy}, the LMC is the second-closest galaxy to the MW. It has a stellar disc radius of $\sim$14 kpc \citep{2010saha, 2019nidever} and a mass of 1.8 $\times$ 10$^{11}$M$_{\odot}$ \citep{2024watkins,2024nikolay}. This is about one-tenth the mass of the MW, making it the fourth most massive galaxy in the Local Group \citep{2012McConnachie}.

The LMC has a non-axisymmetric spiral structure with a flared disc \citep{2002vandermarel, 2022ripepi}. It features an off-centre bar \citep{2001vandermarel, 2012besla} that is warped, meaning different parts of the bar are tilted at various angles relative to the galaxy's plane \citep{2003annapurni, 2009subramanian}. The non-axisymmetric structure of the LMC is attributed to its interaction with the SMC and the MW, which also led to the warping of its stellar disc in the outer regions (at $\sim$2.5 kpc, \citealt{2002olsen}; $\sim$5.5 kpc, \citealt{2018choi}). The LMC is rich in gas and dust, which facilitates active star formation. The LMC's most intense period of star formation occurred between 0.5 and 4 Gyr ago \citep{2021mazzi}. Studies have shown that the star formation history of the LMC is closely synchronised with that of the SMC, suggesting that tidal interactions between the two galaxies play a significant role in their evolution \citep{2022massana}.

The LMC's close proximity makes it an ideal probe for detailed kinematical studies to understand the dynamic effects of galaxy interactions. Early kinematic studies of the LMC used optical spectroscopy of a few stars to measure line-of-sight velocities, revealing differential rotation in the outer regions \citep[][and references therein]{1961feast}. Radio observations of neutral hydrogen subsequently confirmed this rotational behaviour \citep{1984rohlfs, 1998kim}. Later on, different studies employed diverse tracers, including star clusters, nebulae, {H\,\sc{ii}} regions, and different stellar populations, for kinematical analysis \citep[][and references therein]{2009vandermarel}. \citet{2002vandermarel} developed velocity equations for extended galaxies like the LMC and used them to fit kinematic data from carbon stars. They identified the stellar dynamical centre, coinciding with that of the bar and outer isophote but offset from the {H\,\sc{i}} centre. The derived rotation curve and mass estimate indicated the presence of a dark halo. Additionally, the line-of-sight velocity dispersion analysis suggested a thick disc, consistent with simulations. 

A proper motion study using the Hubble Space Telescope (HST) confirmed the clockwise rotation of stars in the LMC disc plane \citep{2006kallivayalil, 2013kallivayalil} and suggested it is currently on its first in-fall to the MW \citep{2007besla}. \citet{2014vandermarel} conducted the first comprehensive kinematic analysis of the LMC using 3D velocities, combining HST proper motions with existing radial velocity data to derive its rotation curve. The dynamical centre obtained from the proper motion is surprisingly offset from the line-of-sight velocity centre \citep{2002vandermarel}, but it is closer to the {H\,\sc{i}} centre. 

A thorough study of the structural and kinematic properties of the LMC was carried out using data from the second and early third Gaia data releases \citep[DR2 and eDR3;][hereafter G21]{2018gaia,2021GaiaMC}. The authors were the first to create velocity maps in the in-plane radial and tangential directions within the LMC disc plane for different stellar populations. They determined the kinematic centre of the LMC, which is closer to the photometric centre and offset from the {H\,\sc{i}} centre, similar to the findings of \citet{2002vandermarel}. Additionally, they examined the kinematic similarities and differences between young and old stellar populations within the LMC, confirming that the older population is in a kinematically hot disc, while the younger stars are in a cold disc. \citet{2022choi} utilised proper motion data from Gaia eDR3 to determine the impact parameter for the most recent direct collision between the LMC and SMC (see also \citet{2018zivick}), while \citet{2023arranz} showed that the kinematics of the inner disc is primarily influenced by the bar and can be analysed independently of line-of-sight velocity information. Furthermore, \citet{2024dhanush} used the Gaia DR3 dataset to construct kinematical models of the LMC disc based on star clusters and field stars, and reported the differences in the parameters derived for the two components. \citet{2024arranz} analysed the kinematics of the LMC bar using Gaia DR3 data and kinematic modelling techniques, concluding that the bar exhibits a stable structure despite the significant dynamical impact of the LMC$-$SMC interaction, while \citet{2024rathore} used the completeness-corrected Gaia DR3 dataset for red clump stars to demonstrate that the LMC has been a barred galaxy for over 100 Myr and has experienced significant structural changes due to a recent direct collision with the SMC approximately 100 Myr ago. 
 
A proper motion study of the LMC exploiting data from the VISTA survey of the Magellanic Clouds system \citep[VMC;][]{2011cioni} was carried out by \citet[hereafter FN22]{2022florian} and \citet[hereafter TS22]{2022thomas}. In \citetalias{2022florian}, the focus was on the inner parts of the LMC, where they discovered the first observational evidence of elongated orbits within the bar of the LMC. The dynamical centre identified in this study is closer to the {H\,\sc{i}} centre, which is consistent with the findings of \citet{2014vandermarel}. They also derived the galaxy's rotation curve, obtaining results similar to those of \citetalias{2021GaiaMC}. \citetalias{2022thomas} conducted a study of the outer regions of the LMC disc using VMC data in conjunction with the Gaia eDR3 dataset. By generating rotational velocity maps, they found evidence of stripped stellar sources from the SMC resulting from the interaction between the LMC and SMC. \citet{2024nikolay} performed dynamical modelling of the LMC, exploiting the VMC dataset for cross-validation. They incorporated a triaxial bar component to derive crucial dynamical parameters, such as the bar's pattern speed and the mass distribution of the galaxy.

The Magellanic Clouds have been dynamically coupled for $\sim$2 billion years \citep{2012diaz} and underwent a significant interaction $\sim$150 million years ago \citep{2005kenji, 2018zivick, 2022choi}. Tidal forces from the MW are expected to eventually merge them with our galaxy \citep{2019cautun}. These interactions reshape the LMC by stripping material from its inner regions (8 deg from the LMC centre) to the outskirts, with its periphery revealing the impact of these gravitational forces. Stellar substructures in the periphery of the Magellanic Clouds have been studied for many years. \citet{2006munoz} identified a foreground population of red clump stars with radial velocities and metallicities similar to those of LMC sources located more than 22 deg from the LMC centre. The existence of a LMC stellar halo was proposed by \citet{2009majewski}, who identified red-giant branch (RGB) sources exhibiting line-of-sight velocities consistent with the LMC, out to a radius of 23 deg from the LMC centre. One of the most prominent substructures in the outskirts of LMC is an arc-like substructure around 13.5 deg north of the LMC centre, discovered by \citet{2016mackey}, using the first data release of images by the Dark Energy Survey \citep[DES;][]{2005abbott}. It stretches more than 10 kpc towards the east in the direction of the Carina Dwarf galaxy and is 1.5 kpc wide. Kinematic analysis of this feature by \citet{2022cullinane} suggests it is strongly influenced by the LMC's infall to the MW, but it has origins in early LMC-SMC interactions. 

Employing data from Gaia DR2, \citet{2019belokurov} identified several structures in the northern and southern regions of the LMC periphery. They simulated the origin of these structures by modelling the infall of the Magellanic Clouds and concluded that both tidal stripping by the MW and interactions between the LMC and SMC are essential for their creation. \citet{2019navarrete} conducted a spectroscopic analysis of the stellar streams identified by \citet{2016belokurov}, and they have suggested that the LMC halo extends significantly farther out than previously estimated. Investigation into the outer substructures of the Magellanic Clouds, utilising near-infrared data from the VISTA Hemisphere Survey \citep[VHS;][]{2013mcmahon} was conducted by \citet{2021dalal}. In their study, they confirmed the presence of previously identified substructures and discovered a new one in the outskirts of the LMC, located to the east. Based on their spectroscopic study on the substructures of the LMC, \citet{2022cullinane} inferred that the northeastern LMC disc has experienced minimal perturbation. Their simulations suggested that the disturbed nature of the western LMC periphery was potentially caused by an LMC$-$SMC interaction, occurring around 400 Myr ago.

Despite extensive studies on the LMC's kinematics and structure, key questions remain, particularly regarding its dynamical centre, which varies across studies. Previous analyses of LMC substructures have mainly relied on colour-magnitude diagrams (CMDs) and model fitting. In this work, we examine the internal kinematics of the LMC by analysing 2D stellar motions within its disc plane using near-infrared data from the VMC survey. For the first time, an additional epoch from the VMC survey is utilised in this work, extending the proper motion baseline from $\sim$2.5 years to $\sim$10 years. This extension significantly improves the precision of proper motion measurements, reducing uncertainties from 6 mas yr$^{-1}$ to 1.5 mas yr$^{-1}$. Leveraging these improved proper motions, we derive the dynamical parameters and construct velocity curves for various stellar populations. Our goal is to identify kinematic evidence of substructures by generating detailed velocity maps and linking inner disc features to larger-scale substructures.

The paper is organised as follows: in Section \ref{sec:data} we provide an overview of the data used in this study. In Section \ref{sec:PM}, we describe the steps followed to derive the proper motion of stars from the VMC survey. We present the theoretical and statistical approaches used to derive the dynamical parameters in Section \ref{sec:velocity_model}. In Section \ref{sec:int_kinematics}, we focus on generating rotation curves and velocity maps within the LMC disc plane and search for kinematic signatures of substructures within the disc. In Section \ref{sec:Dynamical_models}, we use dynamical modelling to study the nature of these signatures and their connection to substructures in LMC periphery. Section \ref{sec:conclusion} summarises the results and discussions of this study. We also discuss potential improvements and outline future research directions.

\section{Data}
\label{sec:data}

The data for this study were obtained from the VMC survey. This is a public survey by the European Southern Observatory (ESO), of the Magellanic system in the near-infrared bands Y, J, and K$_s$, using the Visible and Infrared Survey Telescope for Astronomy \citep[VISTA;][]{2010emerson}. The VISTA telescope is a 4-m class wide-field survey telescope located at ESO's Cerro Paranal Observatory in Chile, equipped with the Vista Infrared Camera \citep[VIRCAM;][]{2006dalton} having a field of view of 1.65 deg in diameter. The VIRCAM consists of 16 Raytheon VIRGO detectors arranged in a 4$\times$4 array, each with a mean pixel size of 0.339\arcsec and gaps in between the detectors. To observe a contiguous area of the sky, the telescope is shifted with small offsets in the x and y planes, producing images called pawprints. The final VISTA image dubbed as a tile is created by stacking six pawprint images together, giving a total area of $\approx$ 1.77 {deg$^2$}. This arrangement results in the overlapping of regions of the sky where 1.50 {deg$^2$} of the tile area is observed at least twice or more, whereas two horizontal strips covering an area of 0.14 {deg$^2$} each at the top and bottom of the tile (for position angle $\phi = 0$) are observed only once. 

\begin{figure}
    \centering
    \includegraphics[height=10cm,width=\columnwidth]{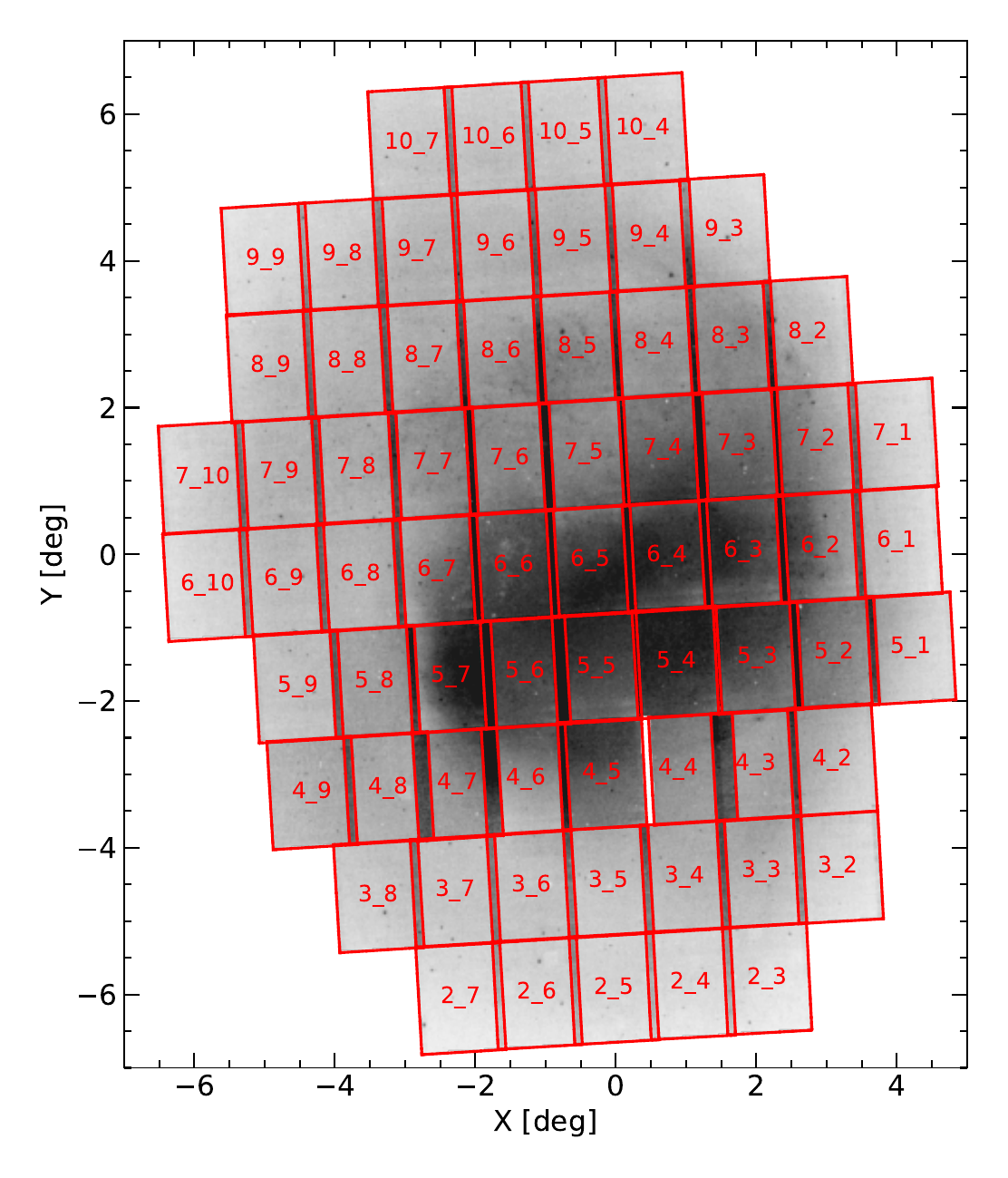}
    \caption{Arrangement of VMC tiles across the LMC field. The coordinates were transformed using zenithal equidistant projection, relative to the dynamical centre of the galaxy, $\alpha = 80.30^{\circ}$, $\delta = -69.27^{\circ}$ (J2000; derived in this study). The VMC sources are plotted in the background, East is to the left and North is to the top.}
    \label{fig:tile}
\end{figure} 

The VMC survey started observations in November 2009 and ended in October 2018; however, additional observations were taken from August 2021 until January 2023 to increase the time baseline for deriving stellar proper motions. The survey takes multi-epoch observations of the Magellanic Clouds, having an average seeing of $\approx$ 0.9\arcsec and an airmass limit of 1.7. A total of 110 VMC tiles were produced for the survey covering a total area of 170 {deg$^2$}, of which 68 tiles, spanning an area of 105 {deg$^2$} were dedicated to the LMC as shown in Figure \ref{fig:tile}. We chose tiles in the K$_s$ band for the proper motion study, to avoid effects of differential atmospheric refraction and as these have the longest time baseline. Most of the tiles in the VMC survey have 13 epochs of good-quality observations available and a time baseline of approximately 2 years. The new observations employed here for the first time, add one more epoch to the tiles, increasing the time baseline to approximately 10 years. For tile LMC 7\_5, the additional epoch was observed as part of a monitoring program for young stars \citep{2020zivkov}, resulting in 18 epochs for this tile over a time baseline of 7 years (see Table \ref{tab:Tiles} for details). The quality of all VMC observations, including the new ones, is described in \citet{2025cioni}, which accompanies the VMC data release 7. The exposure time for a single pawprint image per epoch is 375 s, while two of the epochs have shallow observations with half the exposure time. In a VISTA tile image, the average exposure time per pixel is 750 s, because two pawprint images generally cover most of the tile area.

For this work, individual pawprint images were downloaded from the VISTA Science Archive\footnote{\url{http://horus.roe.ac.uk/vsa}} \citep[VSA;][]{2012cross}. These images were pre-processed by the Cambridge Astronomy Survey Unit (CASU) through the VISTA Data Flow System \citep[VDFS v1.5;][]{2004Irwin,2018Gonz}. The proper motion of stars was calculated using centroids determined by performing point spread function (PSF) photometry on the pawprint images at a detector level. Initially, the pawprint images were unpacked to individual detectors and bad epochs (e.g. those obtained under poor sky conditions or with less than 6 pawprints per tile) were removed. The PSF photometry was performed on individual detector images using a photometric pipeline made by  \citet{2015rubele}, which uses the IRAF\footnote{IRAF is distributed by the National Optical Astronomy Observatories, which is operated by the Association of Universities for Research in Astronomy, Inc. (AURA) under cooperative agreement with the National Science Foundation.} software package; for this study we used an updated version of the code \citep{2021florian}. During the fitting of the PSF profile for isolated stellar sources, the analytic function MOFFAT25 was manually specified in the pipeline, rather than allowing the code to select the function automatically, which was the practice followed in previous studies. This change was prompted by a pipeline issue where the code failed to generate model PSF profiles for all images, producing them for a few when the function selection was automated. Additionally, the look-up table generated by the PSF routine, which contains the fitting parameters, is kept consistent across individual detectors. The total number of images for performing PSF photometry varies depending on the number of epochs available per tile. For instance, in the case of K$_s$ band, with 14 epochs of observations per tile, the total number of images to perform photometry is calculated as 14$\times$6$\times$16 = 1344, where 6 and 16 represent the number of pawprints and detectors, respectively.

We also made use of deep catalogues, produced by \citet{2015rubele} from the epoch-merged multi-band images. These deep catalogues were used for refining the individual epoch catalogues from spurious detections, and moreover, they provide information about colour and morphology essential for identifying background galaxies, which were absent in the individual catalogues.

\section{Determining proper motion of stars}
\label{sec:PM}

For deriving the proper motion of stars in the LMC, we followed the method developed by \citet{2016cioni} and later improved by \citet{2018florian2,2018florian1,2021florian,2022florian}. The proper motion calculations were performed on a per-detector per pawprint level to eliminate systemic offsets while combining them. The individual epoch catalogues per detector obtained from PSF photometry consist of central coordinates of sources in Right Ascension (RA) and Declination (Dec), the x and y pixel coordinates in the detector, and zero-point corrected magnitudes. From the deep-tile catalogues, we selected only sources detected in both the J and K$_s$ bands, and assigned them a unique source ID. Subsequently, the individual epoch catalogues were cross-matched with the deep catalogues, applying a matching radius of 0.2\arcsec. The cross-match removed false detections resulting from PSF photometry and created a consistent set of sources across all epochs. The background galaxies within single-epoch catalogues were identified following criteria based on \citet{2019bell}, and utilised the colour, magnitude, and stellar probability information from the deep dataset. Finally, the individual epoch catalogues were split into two separate sets: one comprising stellar sources and the other consisting of background galaxies. 

\subsection{Common frame of reference}
\label{sec:CFR}
For single-epoch catalogues, there were small offsets in stellar positions attributed to variations in observing conditions and/or shifts in telescope positioning across different epochs. Hence, to measure the intrinsic motion of sources in the LMC, a coordinate transformation of the detector's x and y positions to a common frame of reference was employed using probable LMC sources. The transformations were executed by selecting for each tile a reference epoch to which the observations from all other epochs were transformed. The reference epochs are characterised by the lowest seeing in each tile, ranging between 0.66\arcsec and 0.88\arcsec. The sampling of probable LMC sources for the coordinate transformation was carried out in two steps to obtain the least contaminated sample feasible, as in \citetalias{2022florian}. 

\begin{figure}
    \centering
    \includegraphics[width=\columnwidth]{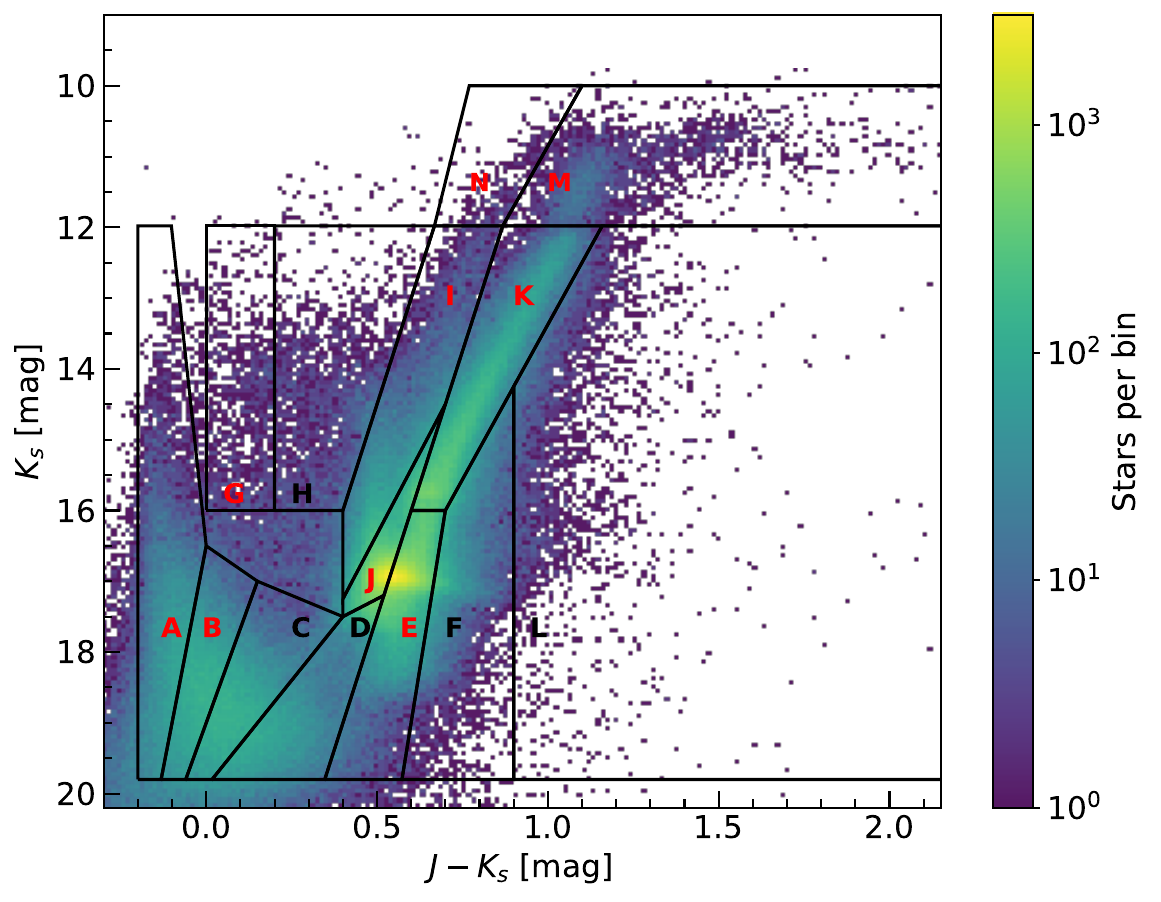}
    \caption{Near-infrared CMD of the Gaia--VMC Deep cross-matched catalogue for the tile LMC 6$\_$5, showcasing probable LMC sources. The distinct stellar regions are depicted with the LMC dominant segments highlighted in red. The background illustrates the stellar density Hess diagram.} 
    \label{fig:cmdgaia}
\end{figure} 

Initially, we selected likely LMC sources from the Gaia eDR3 dataset, adopting the selection criteria outlined in \citetalias{2021GaiaMC}. The Gaia dataset was downloaded from the Gaia data centre at AIP\footnote{\url{https://gaia.aip.de}}, onto which the selection criteria were applied, which resulted in a total of 11,156,431 probable LMC sources with an optical G-band magnitude limit of 20.5 mag. We also tested the probable LMC sources catalogue from \citet{2023arranz}, which was generated using machine learning techniques; however, it did not lead to any significant improvement. Therefore, we chose to use the Gaia eDR3 dataset, which was derived from observable parameters, ensuring a well-defined sample for our study. We then performed a catalogue cross-match using the Astropy Coordinates package\footnote{\citet{astropy:2013, astropy:2018, astropy:2022}}, pairing the Gaia probable LMC sources catalogue and the deep multi-band VMC catalogue employing a matching radius of 0.3\arcsec. The epoch difference between Gaia and VMC was not corrected for, as it was considered negligible ($\approx$ 20 mas for a period of 10 years). Following this step, the majority of bright foreground sources belonging to the MW are eliminated. The next step in refining the sample of probable LMC sources (by removing any remaining foreground sources and background galaxies) involves constructing colour-magnitude diagrams (CMDs) in the near-infrared bands, J $-$ K$_s$ versus K$_s$, for the Gaia--VMC cross-matched sample. The CMD was partitioned into sections representing different stellar populations, derived by \citet{2014cioni}, extending the selection criteria developed by \citet{2000nikolaev} and later modified by \cite{2019dalal}. There are 14 CMD segments comprising categories A, B, and C (young main-sequence stars), D (intermediate-age main sequence and sub-giant stars), E (faint RGB stars), F (MW stars), G, H, I and N (supergiant stars), J (red clump stars), K (bright RGB stars), L (galaxies), and M (asymptotic-giant branch, AGB stars). We selected segments A, B, E, G, I, J, K, M, and N from the CMD, where LMC sources are predominant (see Figure \ref{fig:cmdgaia}).  

\begin{figure}
    \centering
    \includegraphics[width=\columnwidth]{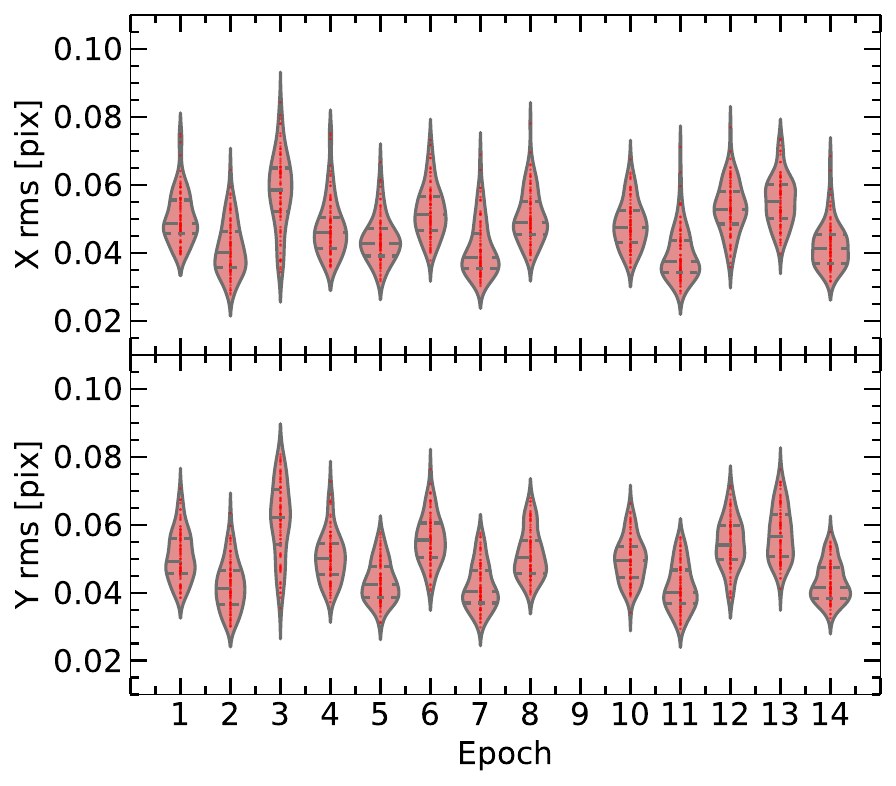}
    \caption{Residual plots for detector x and y positions of tile LMC 2$\_$4, with epoch 9 being the reference epoch, having a seeing of 0.84\arcsec.}
    \label{fig:residual}
\end{figure} 

The subsequent procedure involves transforming the position of sources in the individual epoch catalogues to the reference catalogue, following the steps adopted by \citet{2018smith} (for more details see \citetalias{2022florian}). Afterwards, we compute the root mean square (rms) values for the residuals in both the x and y directions and examine them by creating plots depicting the mean rms values per detector versus individual epochs as shown in Figure \ref{fig:residual}. The residuals across the tiles were less than 0.1 pixels for the outer tiles, where the LMC stellar density is low, and 0.07 pixels for the inner tiles, where the density is higher. However, there were isolated instances where the residuals per detector per epoch exceeded this threshold, particularly for epochs with high seeing conditions. Nevertheless, we noted that as long as the majority of the residual values per detector remained below 0.1 pixels, the derived proper motions were within the expected uncertainty of $\approx$ 1.24 mas yr$^{-1}$. 

\subsection{Relative proper motion calculation}

The single-epoch catalogues, after being aligned to a common frame of reference as discussed above, were now ready for measuring the proper motion of stars. These stars should remain at rest through the epochs following the transformation and any change in their position indicates intrinsic motion around the LMC centre and measurement errors. For the proper motion calculation, we identified stars per detector consistently detected across all epochs, thereby enhancing the number of available positional measurements. For each source, we generated scatter plots of epoch versus position separately for the x and y coordinates and performed a linear least-squares fit of the points. We employed Python's least\_squares function from the Scipy module\footnote{\citet{scipy}\label{fn:scipy}} for fitting, configuring the loss function to linear, which measures how well the model parameters fit the data. This choice of a linear loss function produced optimal results in our analysis \citepalias[see also][]{2022florian}. The slope of the regression yielded the relative proper motion in pixels per day, following the transformations carried out with probable LMC sources. The relative proper motion values were then converted to angular units of milliarcseconds per year (mas yr$^{-1}$) by utilising the pixel scale and orientation of the reference epoch image. We adopted the conventional notation utilising $\mu_W = -\mu_\alpha \cos(\delta)$ for RA (the proper motion in western direction) and $\mu_N = \mu_\delta$ for Dec (the proper motion in northern direction). 

\begin{figure}
    \centering
    \includegraphics[height=12cm,width=\columnwidth]{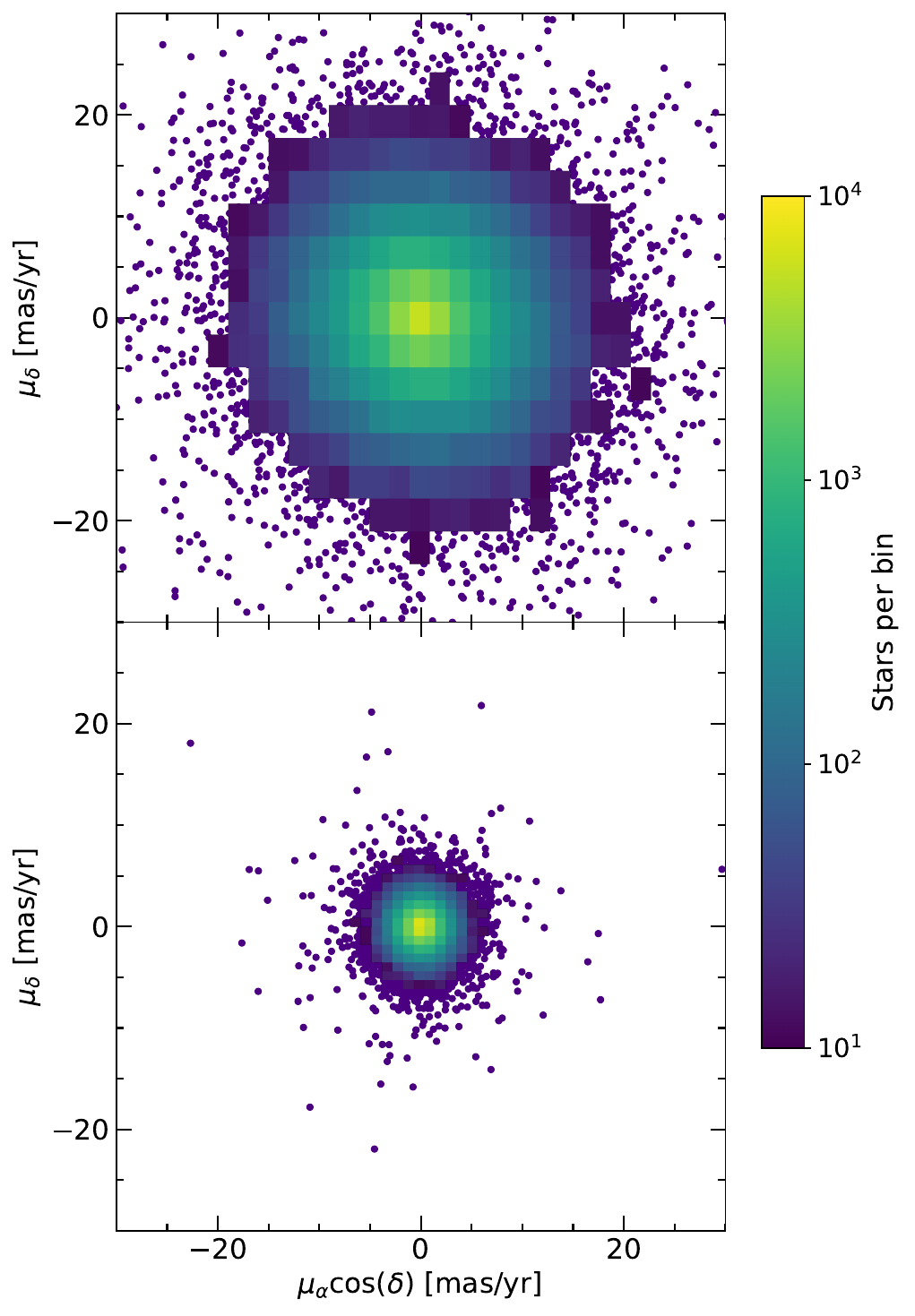}
    \caption{Comparison of relative proper motion calculations for tile LMC 2$\_$4 based on the number of epochs available. The data was binned into 2500 bins, and a threshold was applied based on the stellar counts per bin: only bins with more than 10 stars were included in the density plot. Top: Relative proper motion derived from 13 epochs. Bottom: Relative proper motion corresponding to 14 epochs, including the additional epoch taken in 2023. The extension of one more epoch has increased the time baseline from 3.1 to 9.3 years, improving the precision of the proper motion from 5.7 to 1.5 mas yr$^{-1}$, for this tile.} 
    \label{fig:pm-relative}
\end{figure} 

The proper motion calculations were conducted for a total of 10,214,718 sources within the LMC field. However, due to the overlapping of pawprints resulting in multiple observations of sources, we have identified 5,253,104 unique sources with proper motion measurements, of which about 3,161,000 were used by \citetalias{2022florian} in the inner LMC. In the outer LMC, \citetalias{2022thomas} used a sample of $\sim$2.6 million sources. However, in that work, coordinates were obtained from VDFS, which is based on aperture photometry, while we use PSF photometry. Compared to \citetalias{2022florian} and \citetalias{2022thomas}, this study incorporated the extra epoch observations obtained until January 2023. As a result, the standard deviation per tile of the proper motion decreased from approximately 6 mas yr$^{-1}$ to 1.5 mas yr$^{-1}$ in both $\mu_W$ and $\mu_N$ directions as shown in the density scatter plot in Figure \ref{fig:pm-relative}. 

\subsection{Absolute proper motion}

The alignment of sources to a reference frame utilising exclusively LMC sources only corrects for the observational and instrumental discrepancies. However, to ensure the proper motion values are calibrated to an absolute scale, i.e. heliocentric motion, further conversion is necessary. Before calibrating the relative proper motion catalogues to an absolute scale, it was essential to filter out probable MW sources that may still be present in the catalogue, despite the previous filtering steps, alongside the LMC sources. To accomplish this, we implemented a two-step filtering process. Initially, the VMC proper motion catalogues underwent cross-matching with the Gaia MW catalogue, created by excluding LMC sources from the comprehensive Gaia dataset, using the probable LMC sources catalogue we obtained earlier from Gaia (see Section \ref{sec:CFR}). Afterwards, we utilised the CMD regions, as discussed in section \ref{sec:CFR}, to selectively retain sources predominantly associated with the LMC. As a result of these filtering steps, the total count of unique LMC sources with relative proper motion measurements was reduced to 5,125,009 sources, which represents the most refined sample possible.

To translate relative proper motions into absolute scale, we utilised likely LMC sources from the Gaia catalogue, which was already calibrated to the heliocentric system. Gaia sources were filtered to include only those with renormalised unit weight error (ruwe) $\leq$ 1.4 and proper motion errors less than 0.3 mas yr$^{-1}$. Afterwards, we cross-matched the Gaia and VMC catalogues. The zero points were determined by calculating the difference between the two proper motion measurements and subsequently applied to the VMC proper motion values. However, due to the intrinsic World Coordinate System (WCS) error in the VMC dataset, the proper motion for individual sources was less reliable, but was consistent for a binned dataset \citepalias[see also][]{2022florian}. Finally, we found the median absolute proper motion for LMC sources in $\mu_W$ and $\mu_N$ to be $-$1.85 mas yr$^{-1}$ and 0.34 mas yr$^{-1}$, respectively, with corresponding standard deviations of 1.70 mas yr$^{-1}$ and 1.75 mas yr$^{-1}$.

\begin{figure*}
    \centering
    \includegraphics[width=\columnwidth]{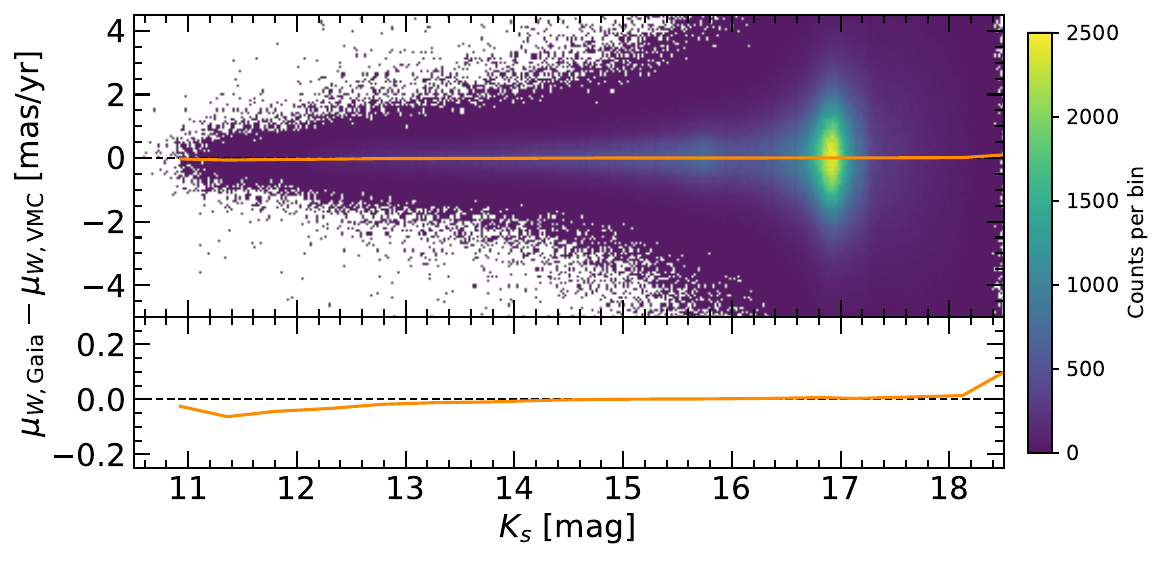}
    \vspace{0.5em}
    \includegraphics[width=\columnwidth]{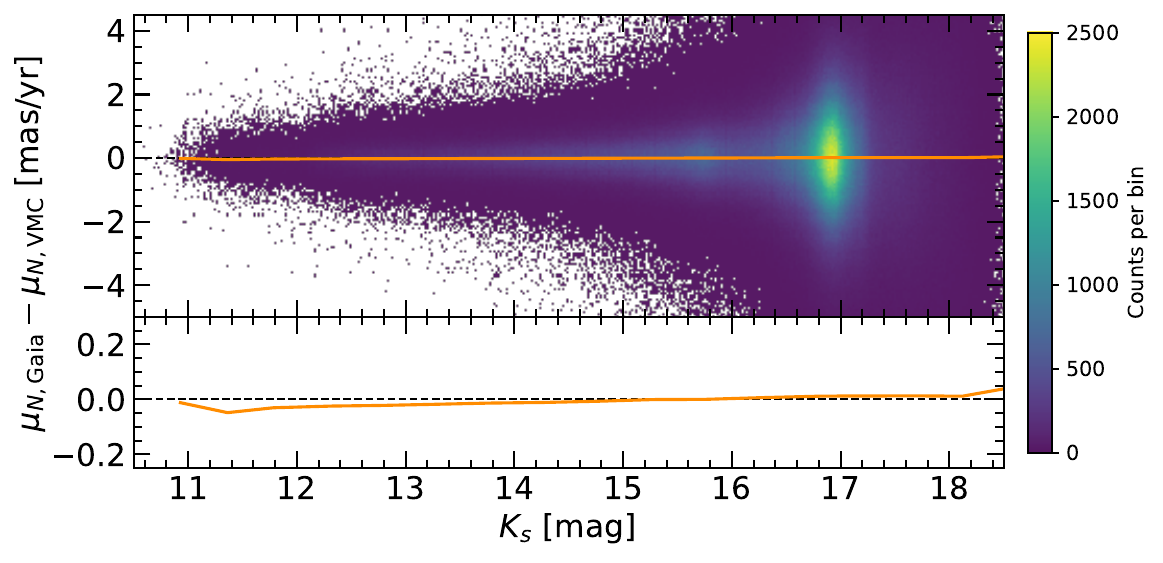}
    \caption{Difference between absolute proper motion measurements for common  sources in VMC and Gaia datasets, shown separately for $\mu_{W}$ (left) and $\mu_{N}$ (right) directions, as a function of K$_{\mathrm{s}}$ magnitude. The orange horizontal lines represent the median difference in 0.5 mag bins. The lower panels provide a zoomed-in view of the median differences to highlight any subtle trends.}
    \label{fig:pm_absolute_comparison}
\end{figure*} 

A comparison of the proper motion measurements for individual stars between the VMC and Gaia eDR3 datasets was performed to validate the reliability and accuracy of the VMC proper motion values. For the LMC sources in common between the two datasets, the median proper motions in the Gaia data were found to be $-$1.83 mas yr$^{-1}$ in $\mu_W$ and 0.37 mas yr$^{-1}$ in $\mu_N$, with standard deviations of 0.35 mas yr$^{-1}$ and 0.48 mas yr$^{-1}$, respectively. The Gaia dataset exhibits a tighter concentration with smaller standard deviations, indicating higher precision compared to the VMC dataset. However, the median difference in proper motion values for individual stars between the two datasets is notably smaller, with a median difference of 0.0006 mas yr$^{-1}$ in the western direction and 0.001 mas yr$^{-1}$ in the northern direction, as shown in Figure \ref{fig:pm_absolute_comparison}. These values are also significantly smaller than those obtained using VMC proper motions in \citetalias{2022florian} (see their Figure 5), which reported offsets of 0.003 mas yr$^{-1}$ in the western direction and less than 0.001 mas yr$^{-1}$ in the northern direction. This indicates an excellent agreement between Gaia and VMC measurements for the same sources. In this work, we will apply data binning to reduce the impact of lower-precision measurements, and the results presented in the following sections are based on this approach.

\section{Modelling of the data}
\label{sec:velocity_model}

\subsection{Velocity field model}

To investigate the internal kinematics of the LMC using absolute proper motion measurements, we adopted the velocity field formalism established by \citet{2002vandermarel}. While \citetalias{2022florian} previously applied this model to the VMC dataset, their analysis was limited to the central 3 kpc. In contrast, our study extends the modelling out to approximately 6 kpc from the LMC centre.

The velocity field assumes a flat, rotating disc galaxy with a large angular extent, as is characteristic of the LMC due to its proximity to the Milky Way \citep{2010saha, 2014vandermarel}. Specifically, the VMC survey covers an angular area of about 12 degrees on the sky in the LMC field (see Figure \ref{fig:tile}). The model accounts for three main components: (1) the systemic motion of the centre of mass (COM), (2) internal stellar rotation within the disc, and (3) precession and nutation effects due to tidal interactions. However, the latter have been shown to be negligible with high-precision HST data \citep{2014vandermarel}. 

The full model includes seven dynamical parameters. Three define the disc’s orientation: the sky coordinates of the dynamical centre $(\alpha_0, \delta_0)$, the inclination angle $i$ (between the sky and disc planes), and the position angle of the line of nodes $\Theta$ (measured North to East). The remaining four describe the LMC’s kinematics: the COM proper motions $(\mu_{W,0}, \mu_{N,0})$, systemic line-of-sight velocity $v_{sys}$, distance to the COM $D_0$, and the rotational velocity profile $V(R)$. The rotation curve was parametrised as:

\begin{equation}
V(R) = V_0 \left[1+\left( \frac{R_0}{R} \right)^{\eta} \right]^{-1/\eta}
\label{eq:rot_curve}
\end{equation}

\hspace{-0.58cm}This functional form, used in both \citetalias{2021GaiaMC} and \citetalias{2022florian}, describes a velocity profile that rises linearly with radius up to $R_0$, beyond which it flattens at $V_0$. The curve’s shape is governed by the parameter $\eta$. All three—$V_0$, $R_0$, and $\eta$—are treated as free parameters in the fitting process.

\subsection{Data fitting}
\label{sec:MCMC}

The dynamical parameters of the LMC were derived by fitting the observed proper motions to the transverse velocity model described above. The dataset was binned into a 150$\times$150 grid, with each element covering 19.44 arcmin$^2$ on the sky. Only grid elements containing $\geq$100 stars were included. For each, median values of ($\alpha$, $\delta$), $\mu_W$, and $\mu_N$ were computed, along with uncertainties as the error of the mean.

Following \citetalias{2022florian}, we employed a Bayesian inference framework to estimate the parameters and their uncertainties. Depending on the stellar population, flat or Gaussian priors were adopted. The log-likelihood function used was:

\begin{multline}
    \ln \mathcal{L} = -0.5\left( \sum_{i=1}^{n} \ln (2\pi\sigma_{\mathrm{W},\mathrm{i}}^2) +\frac{(\mu_{\mathrm{W},\mathrm{i}} - \mu_{\mathrm{W},\mathrm{mod},\mathrm{i}})^2} {\sigma_{\mathrm{W},\mathrm{i}}^2} \right.\\
    + \left.\sum_{i=1}^{n} \ln (2\pi\sigma_{\mathrm{N},\mathrm{i}}^2) +\frac{(\mu_{\mathrm{N},\mathrm{i}} - \mu_{\mathrm{N},\mathrm{mod},\mathrm{i}})^2} {\sigma_{\mathrm{N},\mathrm{i}}^2} \right),
\end{multline}

\noindent where $\sigma_W$ and $\sigma_N$ are the proper motion uncertainties in RA and Dec, respectively, and $i$ denotes the grid cell. Model proper motions ($\mu_{W,\mathrm{mod}}, \mu_{N,\mathrm{mod}}$) were computed using Equation 7 of \citet{2002vandermarel}.

We used the Affine Invariant MCMC Ensemble Sampler \citep{2010MCMC} implemented via the \texttt{emcee} package \citep{2013emcee}, with 200 walkers over 2000 steps. Posterior distributions were estimated from the final 25\% of each chain. In the transverse velocity model, the line-of-sight distance, $D_0$ and the line-of-sight velocity, $v_{sys}$ were kept as constants at 49.9 kpc \citep[][this distance was chosen because it is more robust and accounts for distances derived from diverse stellar populations]{2014degrijs} and 262.2 km s$^{-1}$ \citep{2002vandermarel}, respectively, and the rest were set as free parameters. We applied priors for the free parameters based on values obtained by \citetalias{2022florian}.

\begin{figure*}
    \centering
    \includegraphics[height=16cm,width=16cm]{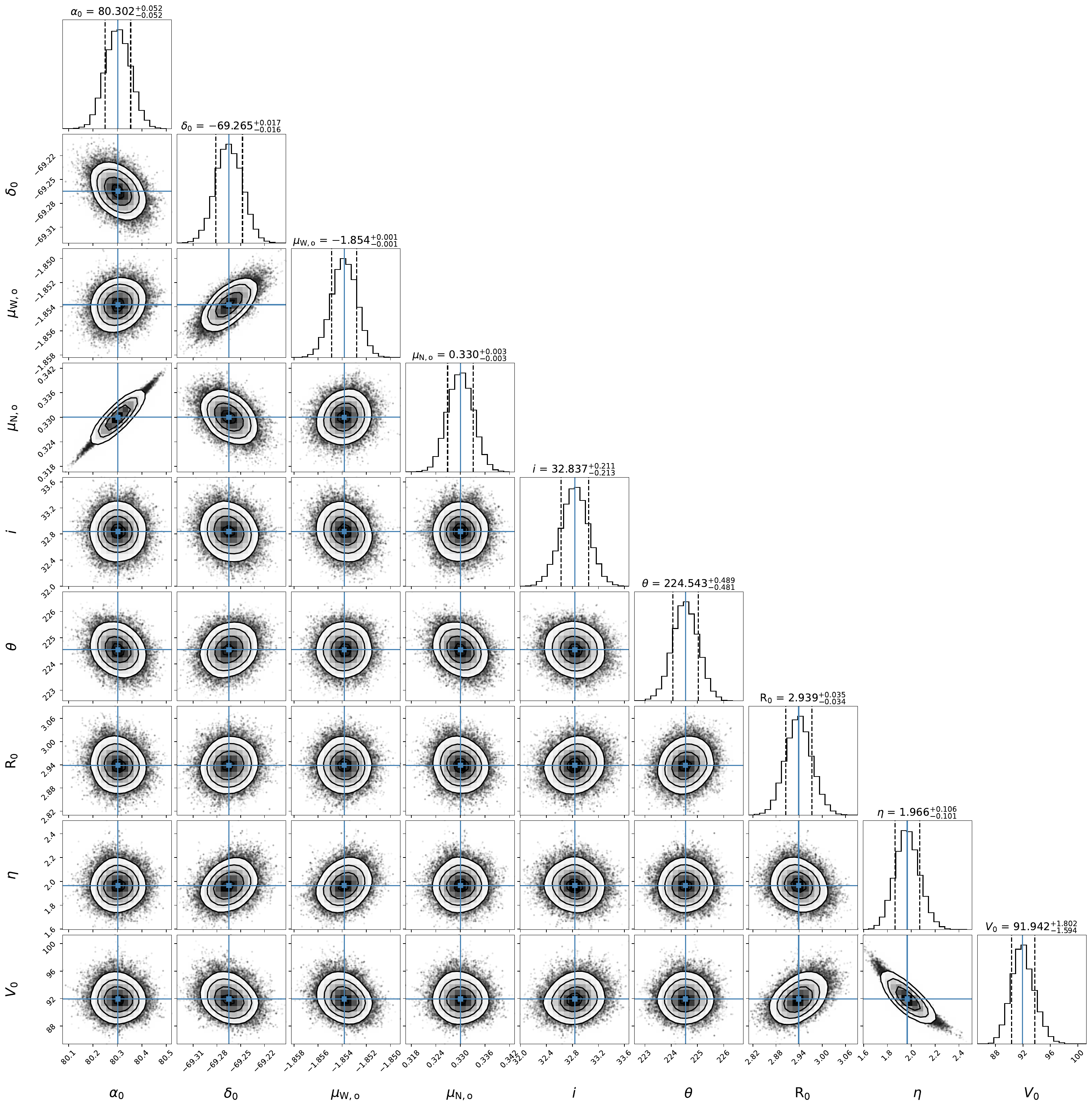}
    \caption{Corner plot showing the joint and marginalised distribution of the nine dynamical parameters for the velocity field model of the LMC obtained for the entire sample. The parameters include the position of the dynamical centre ($\alpha_0,\delta_0$), the proper motion of the COM ($\mu_{W,0}, \mu_{N,0}$), the inclination angle ($i$), the position angle of the line of nodes ($\Theta$). The rotation curve is characterised by the scale radius ($R_0$), the exponential coefficient ($\eta$) and the constant velocity ($V_0$).  Here, $\theta$ represents the angle measured counter-clockwise from the West direction towards the North, while the position angle $\Theta$ is defined as $\theta - 90^\circ$, measured counter-clockwise from North to East.}
    \label{fig:cornerAll}
\end{figure*}

The optimal dynamical parameters were derived for the complete dataset as well as for two subsets categorised by stellar age (see Figure \ref{fig:cornerAll} and Table \ref{tab:param}). The subsets were created based on the position of sources within the CMD as detailed in section \ref{sec:CFR}. Sources falling within sections A, B, G, and N were categorised as the young stellar population ($\lesssim$ 0.5 Gyr), while those within sections E, K, M, and J were considered part of the old population ($\gtrsim$ 1 Gyr). The proper motion dataset of the young population was binned into 70$\times$70 bins due to their lower number density for the fitting. During the MCMC sampling, a uniform flat prior distribution was applied to both the whole sample and the old population. In contrast, a Gaussian prior distribution was employed for the young population because their lower number density required restricting the sample space to produce meaningful results. For the Gaussian prior distribution, a sigma standard deviation of 0.3 was adopted for all parameters except for $V_0$, which was allowed a sigma standard deviation of 3.0 in order for it to have a larger sample space. Additionally, for the young population, the proper motion of the COM was fixed to that of the entire sample, as the bulk motion does not vary significantly for different stellar populations. This resulted in only seven free parameters for the fitting process.

In this study, we determined the dynamical centre of the LMC to be ($\alpha_0,\delta_0 = 80.30^\circ \substack{+0.05 \\ -0.05}, -69.27^\circ \substack{+0.02 \\ -0.02}$, epoch J2000), which falls within the error limits of recent values reported in the literature. Specifically, our results are consistent with those of other photometric studies such as \citetalias{2022florian}, who studied the proper motion of LMC sources using inner VMC tiles, and \citet{2022choi}, who analysed the proper motion of red clump stars in the LMC using the Gaia eDR3 dataset. Additionally, our results are in good agreement with the dynamical centre derived by \citet{2024nikolay} through fitting a 3D Jeans dynamical model to the Gaia DR3 dataset. The geometric parameters, namely the inclination angle ($i = 32.8^\circ \substack{+0.2 \\ -0.2}$) and the position angle of the line-of-nodes ($\Theta = 134.5^\circ \substack{+0.5 \\ -0.5}$), are consistent with literature values for comprehensive LMC stellar populations (\citealt{2020wan}; \citetalias{2021GaiaMC}). Similarly, the systemic proper motion ($\mu_{W,0}, \mu_{N,0}$) values are in good agreement with those reported in the literature (\citetalias{2021GaiaMC}; \citealt{2022choi}; \citetalias{2022florian}) and the correlation between systemic motion and dynamic centre is clearly visible in the corner plot from the MCMC sampling. The dynamical parameters reveal both consistencies and discrepancies across different stellar populations. Notably, the position angle of the younger population deviates from that of the older population, suggesting that younger stars may occupy a distinct disc plane. This interpretation is further supported by a shift in the dynamical centre along the declination axis. Moreover, the younger population displays a smaller scale radius, indicative of a more centrally concentrated mass distribution relative to the older population. The variation in rotation velocity between the two populations is discussed in detail in Section \ref{sec:vel_curves}.

\begin{table}[h]\small
\centering
\caption{Dynamical parameters and their associated uncertainties obtained from the MCMC sampling.}
\begin{tabular}{crrr}
\hline\hline
\noalign{\smallskip}
Parameters & Entire Sample & Old population & Young Population \\ 
\noalign{\smallskip}
\hline
\noalign{\smallskip}
$\alpha_0$ \\ (deg)  & $80.30 \substack{+0.05 \\ -0.05}$ & $80.26 \substack{+0.06 \\ -0.06}$ & $80.20 \substack{+0.06 \\ -0.06}$ \\
\noalign{\smallskip}
$\delta_0$ \\ (deg) & $-69.27 \substack{+0.02 \\ -0.02}$ & $-69.21 \substack{+0.02 \\ -0.02}$ & $-69.64 \substack{+0.04 \\ -0.04}$ \\
\noalign{\smallskip}
$\mu_{W,0}$ \\ $(\mathrm{mas \,yr^{-1}})$  & $-1.854 \substack{+0.001 \\ -0.001}$ &  $-1.854 \substack{+0.001 \\ -0.001}$ & \multicolumn{1}{c}{\textendash} \\
\noalign{\smallskip}
$\mu_{N,0}$ \\ $(\mathrm{mas \,yr^{-1}})$ & $0.330 \substack{+0.003 \\ -0.003}$ & $0.328 \substack{+0.003 \\ -0.003}$ & \multicolumn{1}{c}{\textendash} \\
\noalign{\smallskip}
$i$ \\ (deg) & $32.8 \substack{+0.2 \\ -0.2}$ & $32.8 \substack{+0.2 \\ -0.2}$ & $32.8 \substack{+0.3 \\ -0.3}$ \\
\noalign{\smallskip}
$\Theta$ \\ (deg) & $134.5 \substack{+0.5 \\ -0.5}$ & $136.5 \substack{+0.5 \\ -0.5}$ & $134.3 \substack{+0.3 \\ -0.3}$ \\
\noalign{\smallskip}
$R_0$ \\ $(\mathrm{kpc})$ & $2.94 \substack{+0.04 \\ -0.03}$ & $3.14 \substack{+0.04 \\ -0.04}$ & $2.47 \substack{+0.16 \\ -0.16}$ \\
\noalign{\smallskip}
\noalign{\smallskip}
$\eta$ & $2.0 \substack{+0.1 \\ -0.1}$ & $2.0 \substack{+0.1 \\ -0.1}$ & $2.1 \substack{+0.2 \\ -0.2}$ \\
\noalign{\smallskip}
$V_0$ \\ $(\mathrm{km \,s^{-1}})$ & $91.94 \substack{+1.80 \\ -1.59}$ & $92.45 \substack{+1.95 \\ -1.72}$ & $94.63 \substack{+2.77 \\ -2.52}$ \\
\noalign{\smallskip}
\hline
\label{tab:param}
\end{tabular}
\tablefoot{The systemic motion of the young population is fixed to the value of the entire sample (see text for details).}
\end{table}

\section{Internal kinematics of LMC}
\label{sec:int_kinematics}

In this study, the internal kinematics of the LMC was examined by deprojecting the velocities from the sky plane to the LMC disc plane. The sky plane is the zenithal equidistant projection of the celestial sphere centred on the LMC dynamical centre \citep[see equation 4 in][]{2001vandermarelcioni}. The LMC disc plane is produced by rotating the x-axis of the sky plane counter-clockwise about the direction perpendicular to the sky plane by an angle $\theta$, which is the position angle, and tilting it by the inclination angle $i$. The de-projection equation for converting velocities in the sky plane to the LMC disc plane was adopted from equations 5, 7, and 11 in \citet{2002vandermarel}, solving for $v_x^{\prime}$ and $v_y^{\prime}$. The de-projection was achieved by employing the dynamical parameters outlined in Table \ref{tab:param} for specific stellar populations. After de-projection, the kinematic analysis of the LMC was conducted by generating velocity curves based on radial distance from the LMC centre, as well as 2D velocity maps in the LMC disc plane. 

\begin{figure*}
    \centering
    \includegraphics[height=6cm,width=\columnwidth]{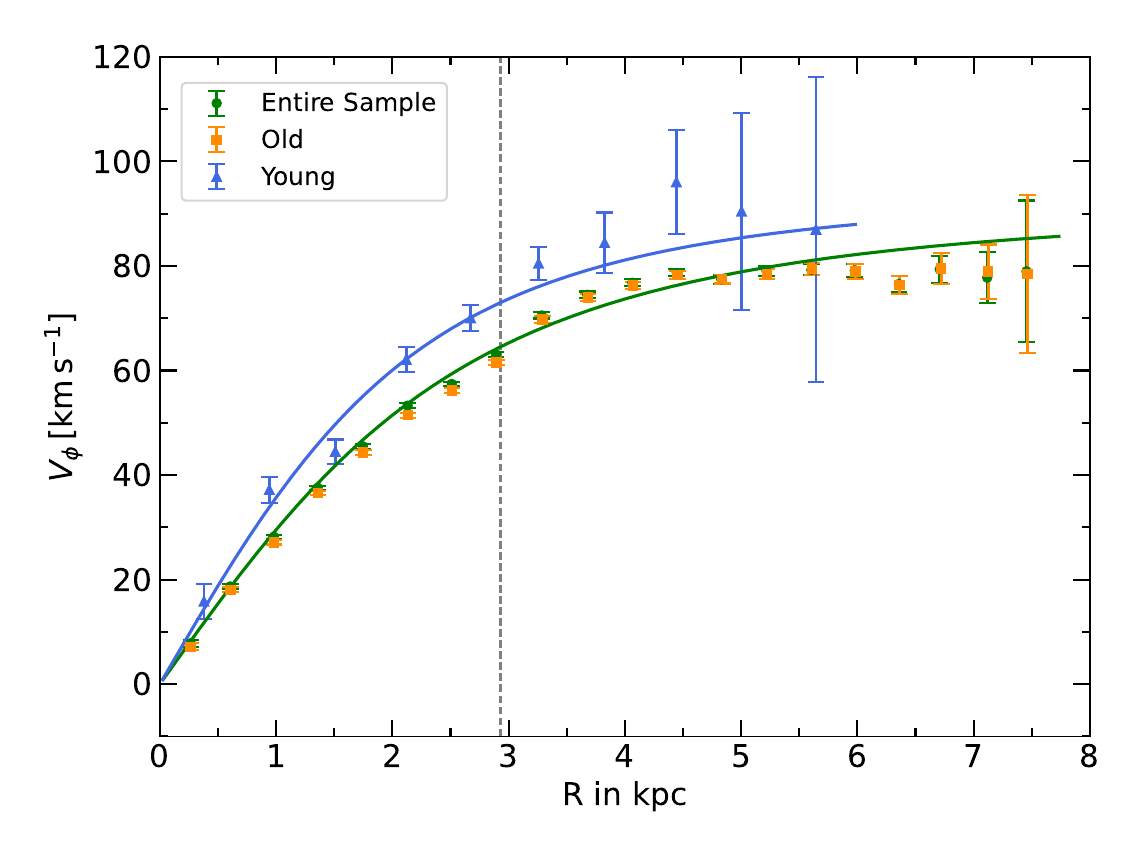}
    \hfill
    \includegraphics[height=6cm,width=\columnwidth]{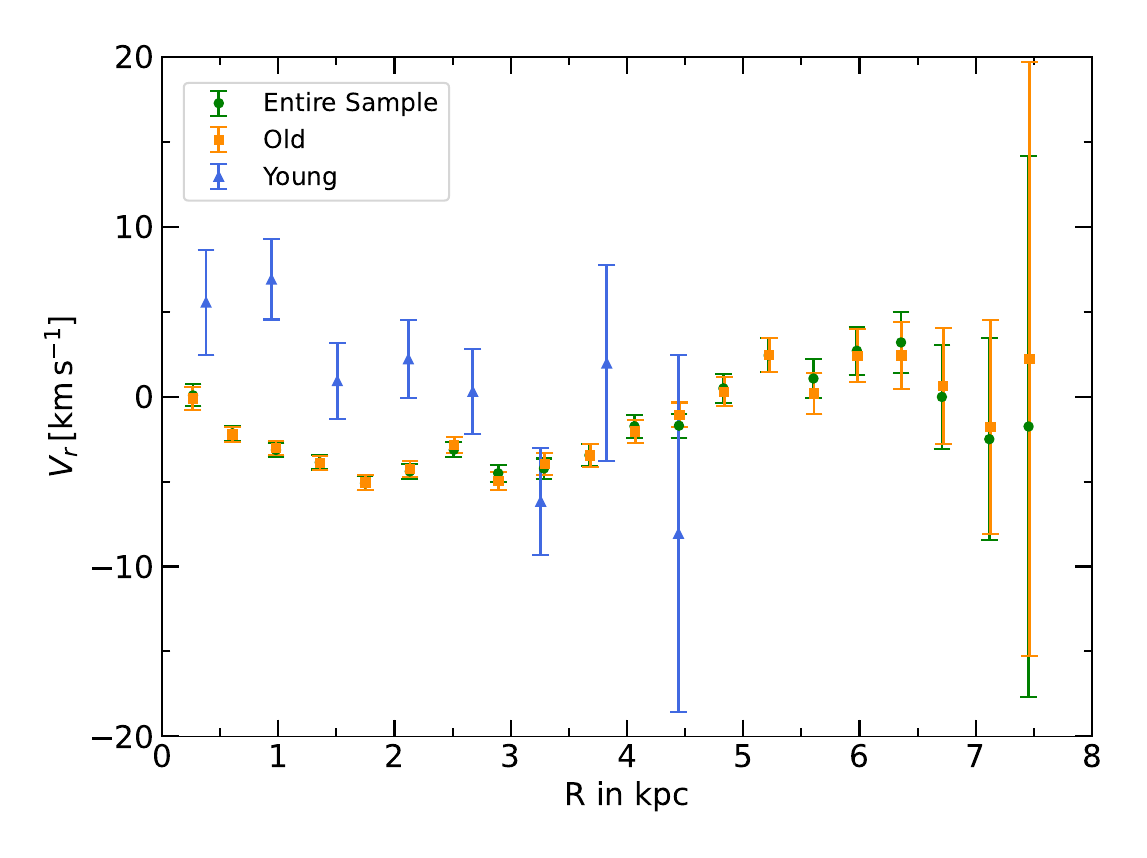}
    \caption{Rotation (left) and radial velocity curves (right) of the LMC for different stellar populations. In the rotation curve (left), the vertical dashed line indicates the estimated $R_0$ value for the entire sample from the MCMC sampling. The green and blue solid lines represent the theoretical rotation curve for the whole sample and the young population, respectively.}
    \label{fig:vel_curve}
\end{figure*} 

\begin{figure}
    \centering
    \includegraphics[width=\columnwidth]{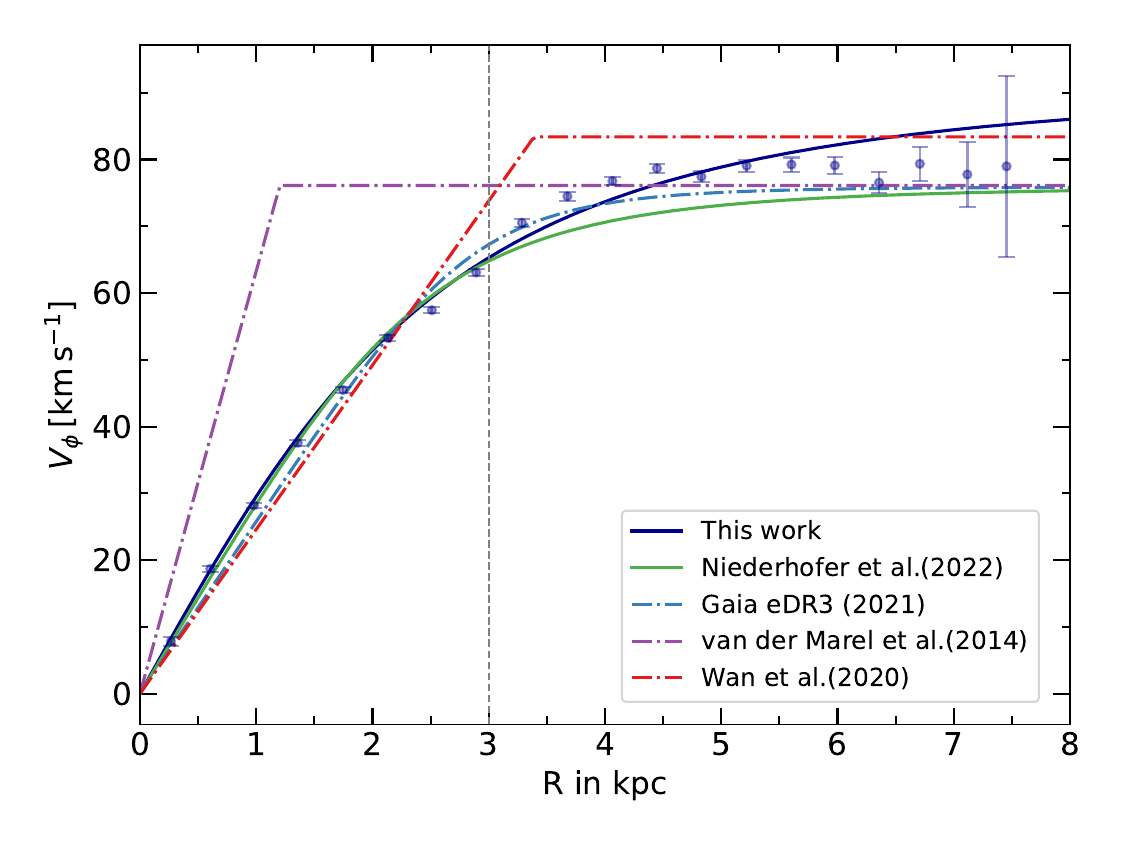}
    \caption{Comparison of rotation curve models with literature results. The dark blue line represents the model used in this work, and the data points are the same as those shown in Figure \ref{fig:vel_curve} for the entire sample. The solid green curve corresponds to the velocity model from \citetalias{2022florian}, and the vertical dashed line indicates their radial coverage. The dash-dotted lines represent models from the literature based on Gaia eDR3 \citepalias{2021GaiaMC}, HST \citep{2014vandermarel}, and SkyMapper DR1 \citep{2020wan}.}
    \label{fig:vel_curve_models}
\end{figure} 

\subsection{Velocity curves}
\label{sec:vel_curves}
To study the velocity curves, the de-projected velocities in the LMC plane were transformed to polar coordinates in radial and azimuth directions. The velocities were then partitioned into 20 radial bins ($\approx 0.4$ kpc per bin) for both the entire sample and the old population, whereas the young population was divided into 10 radial bins ($\approx 0.77$ kpc per bin). The median velocities were calculated for each bin to create the rotation and radial velocity curves shown in Figure \ref{fig:vel_curve}. The velocity curves were plotted for the entire sample, as well as for the young and old populations.

The rotation velocity curve, as expected, rises linearly up to a radius of $R_0 = 2.94$ kpc. This value for the scale radius matches the \citetalias{2021GaiaMC} value for the comprehensive LMC stellar population. The green solid line in Figure \ref{fig:vel_curve}, left panel, represents the theoretical rotation curve derived from equation \ref{eq:rot_curve}, inserting our best-fit parameters, and it aligns well with the observed data. Moreover, the rotation curve of LMC is similar to the rotation curve of other dwarf spiral galaxies \citep{2017karukes}. The rotation curve for the old population is similar to that of the entire sample, as it constitutes the majority of the dataset. In contrast, the young population exhibits a steeper rise, with a scale radius of $R_0 = 2.47$ kpc, and the solid blue line represents its theoretical rotation curve. The asymptotic rotation velocity for the entire sample was estimated to be $V_0 = 91.94$ km s$^{-1}$, which is on the higher end of values reported in the literature \citep{2020wan}. In addition, the asymptotic velocity is higher for the young population compared to the old population in our sample, which indicates asymmetric drift \citepalias[noted by][]{2021GaiaMC}. This is due to the higher velocity dispersion in the older population, likely due to the influence of the bar or previous interactions involving the SMC. While two previous studies have provided evidence of negative rotation velocity in the velocity curve (\citetalias{2021GaiaMC}; \citetalias{2022florian}), this phenomenon was not observed in our dataset. This discrepancy is attributed to our use of dynamical parameters specific to the stellar population while doing the de-projection. In \citetalias{2022florian}, the negative rotation velocities were not seen when they used literature values for deprojecting their young stellar population sample, which initially indicated negative rotation velocities. This suggests that the young population rotates in a disc that is offset from that of the older population, as the primary difference in dynamical parameters between these populations are the position angle of the line of nodes and their dynamical centre (see Section \ref{sec:MCMC}). This offset could be attributed to the interaction between the LMC and the SMC. The interactions likely affected the gas and stars differently, causing the gas to experience a greater shift. As a result, the young stellar population, which formed within this displaced gas, also exhibits an offset in its morphological parameters. Finally, the rotation curve exhibits a bump between 3.5 and 5 kpc, which was present despite adjustments to the dataset binning, followed by a decline beyond 5.5 kpc. This decline in the LMC rotation curve has been noted in previous studies (\citealt{2000alves, 2014vandermarel, 2020wan}; \citetalias{2021GaiaMC}; \citealt{2022choi}). These features may be attributed to the influence of the galactic bar, which induces non-circular stellar orbits in the disc \citep{1986athanassoula}, consistent with the LMC's inherently elliptical nature \citep{2001vandermarelcioni}. However, the decline in the outer regions could also be a consequence of other factors linked to the dark matter distribution of the LMC.

A comparison of the rotation curve models with previous studies is presented in Figure \ref{fig:vel_curve_models}. The rotation curves derived in this work and by \citetalias{2022florian} show good agreement up to a radius of 3 kpc, which marks the radial coverage limit of the latter. Afterwards, the curves deviate, with the one from this work flattening at a higher rotation velocity. This discrepancy arises because the rotation velocity of \citetalias{2022florian}, was fixed to 76 $(\mathrm{km ,s^{-1}})$ based on values from previous studies (\citealt{2014vandermarel}; \citetalias{2021GaiaMC}; \citealt{2022choi}). The rotation curve based on Gaia eDR3 data \citepalias{2021GaiaMC} exhibits slight deviations within 4 kpc, beyond which it flattens to a significantly lower rotation velocity similar to \citetalias{2022florian}. The curve derived from HST observations \citep{2014vandermarel} shows more pronounced discrepancies, although at larger radii it converges with the Gaia eDR3 and \citetalias{2022florian} models. The rotation curve obtained from SkyMapper DR1 \citep{2020wan} follows a similar trend within the inner 2.5 kpc and supports a higher rotation velocity at larger radii, but has a larger scale radius compared to our result.

The radial velocity curve obtained in this study is consistent with the general trends in the literature (\citealt{2020wan}; \citetalias{2021GaiaMC}; \citealt{2023arranz}). The velocity curve initially declines to negative values (approximately down to $-$5 km/s) within a radius of 2.5 kpc from the centre of the LMC, indicating the bulk motion of stars towards the galactic centre following the gravitational potential of the LMC. Analogous to the rotation velocity curve, the old population follows the behaviour of the entire sample. For the young population, the error bars in the median velocities become too large to provide any meaningful results beyond 4.5 kpc. This resulted in the dataset being restricted within this radius, as young stars are primarily located in the inner regions such as the galactic bar or spiral arms \citep{2005kenji}. The radial velocity of the young population follows an opposite trend compared to the old. Within $\sim$3.5 kpc, the old population exhibits a negative radial velocity, while the young population shows a positive velocity. This positive radial velocity may be linked to outward radial gas flow caused by the bar structure in the LMC, leading to star formation around 0.2 Gyr ago \citep{2005kenji}. Beyond this distance, the trend reverses. Notably, the radial velocity of the young population demonstrates a declining trend with increasing radial distance; however, the error bars for the young population are quite high, which complicates the interpretation of these trends. These kinematical differences between young and old populations could be compared with the prediction of existing or future simulations of the LMC dynamics.

\subsection{Velocity maps}

\begin{figure}
    \centering
    \includegraphics[width=\columnwidth]{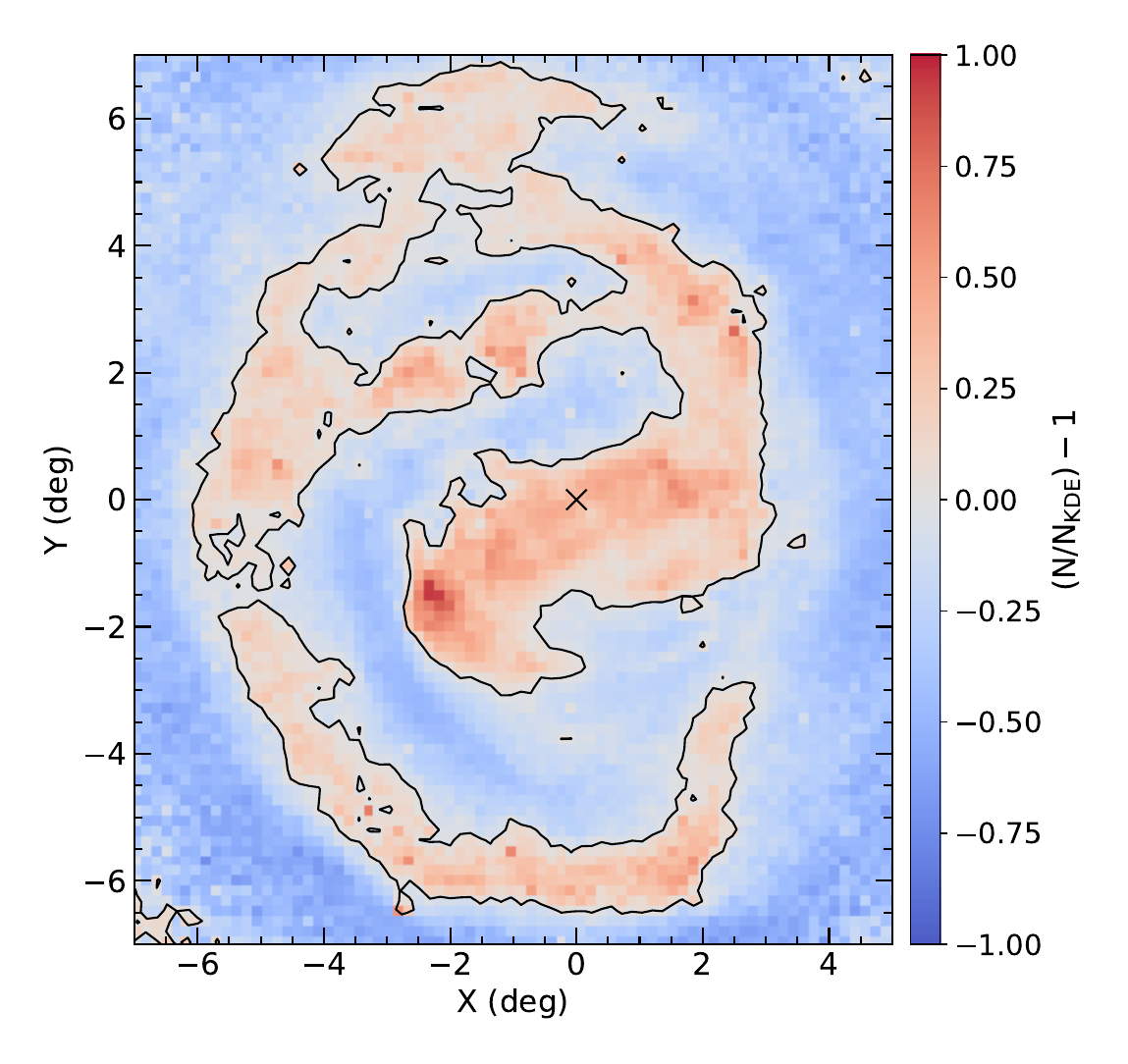}
    \caption{Density contour plot highlighting the off-centre bar and the spiral arm of the LMC projected in the disc plane and aligned with the sky plane. The contour was created using likely LMC sources in the Gaia eDR3 dataset with a KDE bandwidth of 0.4 deg. The dynamical centre derived in this study is indicated by a black cross.} 
    \label{fig:density_contour}
\end{figure} 

Vector diagrams aid in visualising the spatial distribution of velocities across a plane. In this study, we utilised arrow plots and velocity maps to analyse the LMC stellar velocities in the sky plane. Having de-projected the velocities to the LMC plane in the previous section, we subsequently rotated them counter-clockwise by the position angle $\theta$, effectively aligning the velocities to the sky plane \citep{2001vandermarel}. To emphasise the off-centre bar and the spiral arm of the LMC, we employed density contours using probable LMC sources from the Gaia eDR3 dataset (see Section \ref{sec:CFR}), following the methodology outlined by \citet{2023arranz}. Briefly, we computed the Gaussian kernel density estimate (KDE) for the Gaia dataset that was binned into 100$\times$100 bins around 8 deg from the LMC centre. A bandwidth of 0.4 deg was used for the Gaussian kernel function. Subsequently, a box integration was performed on the kernel density estimate for each bin to derive the number density. To accentuate the central bar and the spiral arm, we applied a mask using $(N/N_{KDE}) - 1$ to the binned data, where $N$ denotes the observed number density, while $N_{KDE}$ represents the number density estimated through kernel density estimation. Here, positive values indicate higher density regions, while negative values indicate lower density regions. The density contour highlighting the over-dense regions, projected in the sky plane, is shown in Figure \ref{fig:density_contour}.

\subsubsection{Internal rotation motion}

\begin{figure*}
    \centering
    \includegraphics[width=\columnwidth]{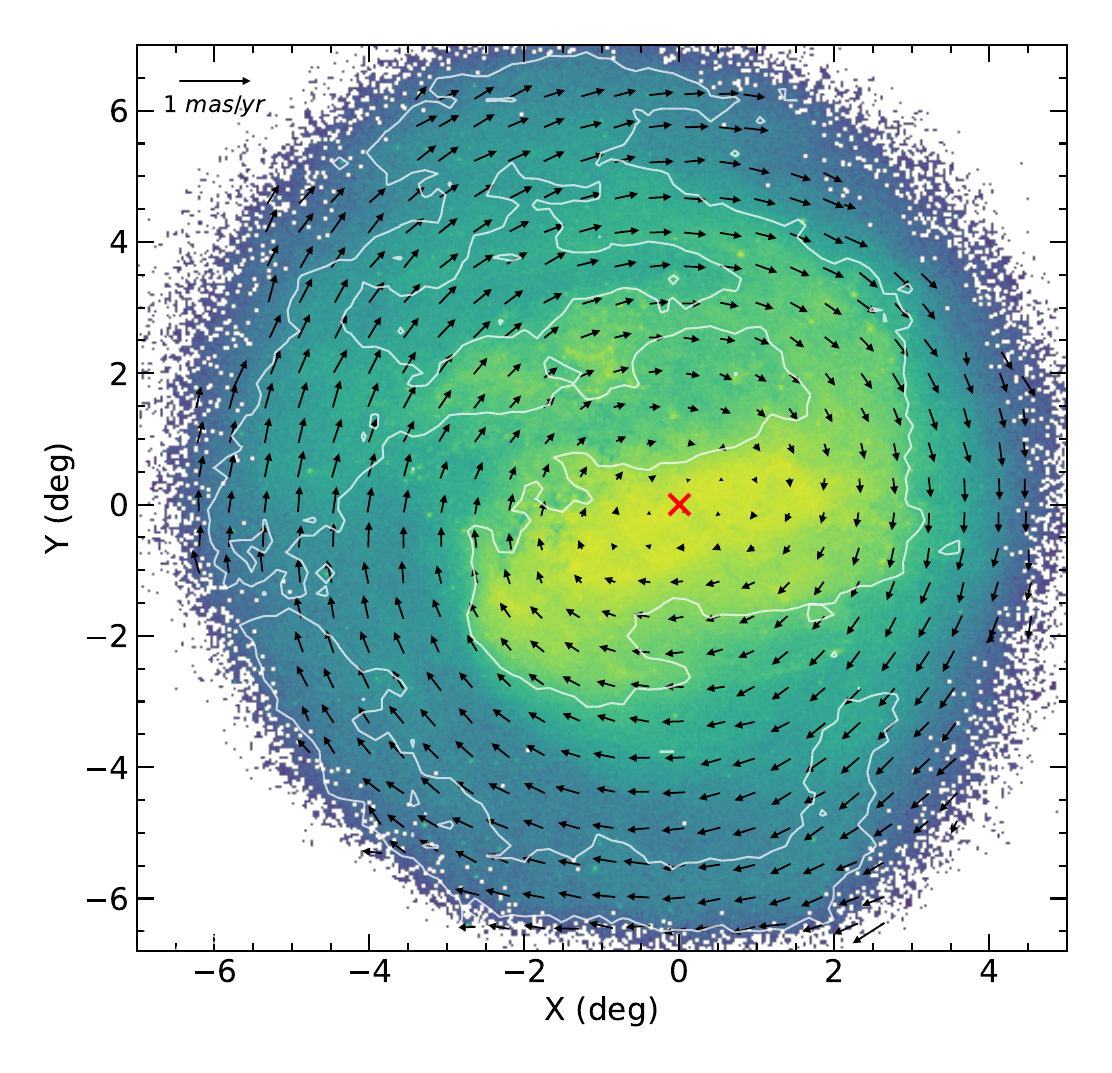}
    \hfill
    \includegraphics[width=\columnwidth]{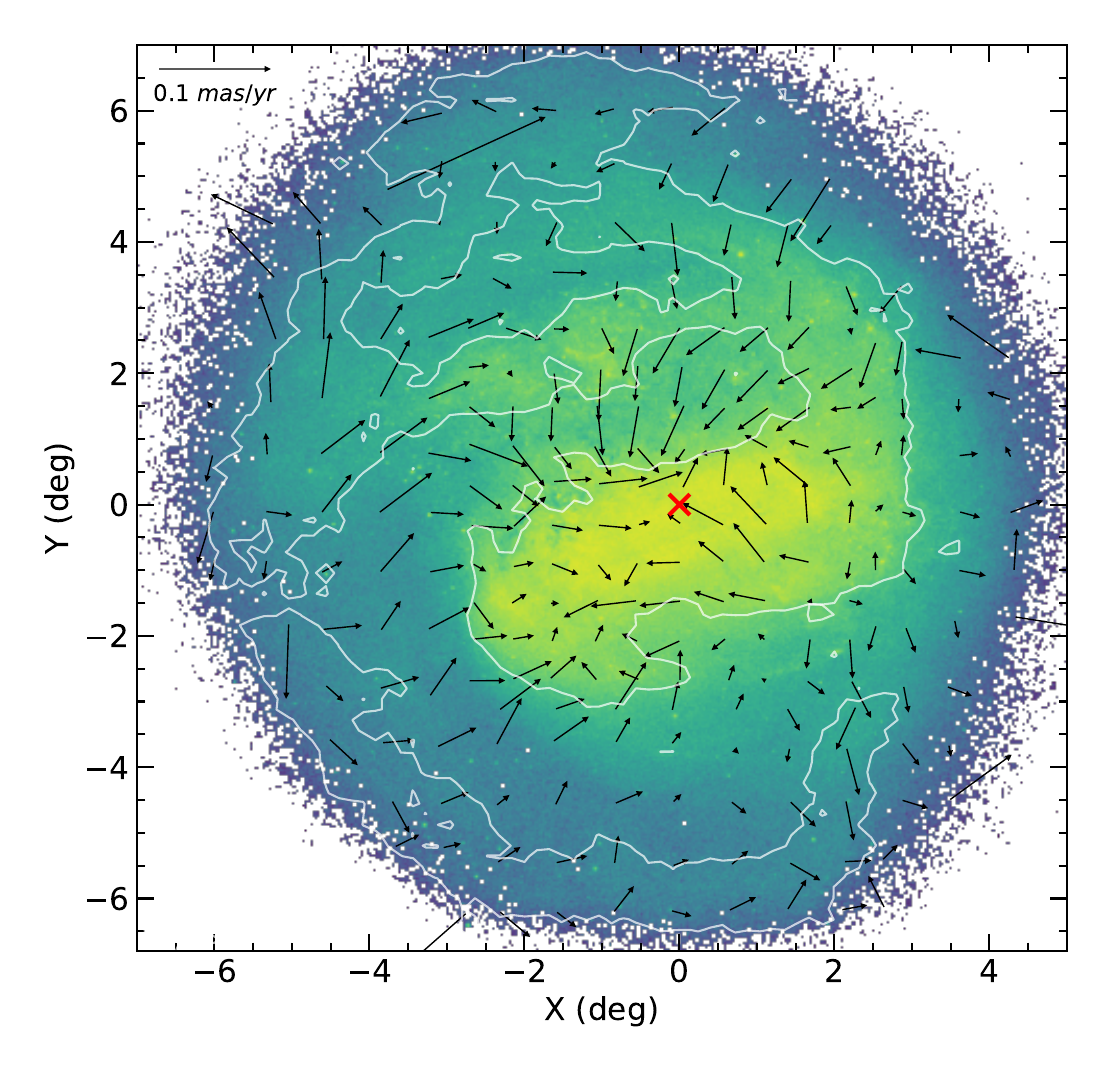}
    \caption{Vector diagram of proper motion in the LMC disc plane aligned with the sky plane. Left: Internal rotation due to the streaming motion of sources in the LMC disc, obtained by subtracting the contribution of the COM motion from the observed proper motion of each object. Right: Residual proper motion (observed proper motion -- model proper motion), encompassing both the COM motion and internal rotational motion. A reference length for the vectors is shown. The background displays the density of likely LMC sources from the Gaia eDR3 dataset. The white contour line highlights the over-density regions, i.e. the bar and spiral arm. The dynamical centre derived for the entire sample in this study is indicated by a red cross. East is to the left and north is to the top.}
    \label{fig:vel_arrow}
\end{figure*} 

The left-hand panel of Figure \ref{fig:vel_arrow} displays the arrow plot showing the internal rotation motion of LMC sources in the sky plane. The background illustrates the number density of likely LMC sources from the Gaia eDR3 dataset, confined within an 8 deg radius from the centre of the LMC. These sources were organised into 1000$\times$1000 bins, where each bin encompasses an area of approximately 63 arcmin$^2$. The bar and the spiral arm were highlighted using the previously derived over-density contour.

To calculate the tangential rotation motion, the contribution from the COM motion was subtracted from the observed proper motion. As already mentioned in Section \ref{sec:velocity_model}, the projection of the COM motion for individual LMC sources changes with viewing angle, and hence they were calculated using equation 13 from \citet{2002vandermarel}, employing the systemic COM proper motion (see Table \ref{tab:param}). The data were binned into 25$\times$25 bins, where each bin corresponds to an area of approximately 0.2 deg$^2$. The median proper motion was computed per bin having a stellar density of more than 100 stars. The stars exhibit a uniform circular motion in the clockwise direction around the centre of the LMC, consistent with theoretical expectations. The vector plot for the old population is identical to the entire sample and showcases a circular motion in the clockwise direction (see Figure \ref{fig:vel_arrow_old}). For the younger population, the sample was limited to within 4 deg of the LMC centre due to their decreasing density outside the over-dense regions. This subset also exhibits clockwise motion (see Figure \ref{fig:vel_arrow_young}). Along the bar, the LMC sources are observed to follow elongated orbits about the bar's major axis. According to galactic bar dynamics, these orbits belong to the x$_1$ family of periodic orbits, which help to stabilise the bar structure (\citealt{1980contopoulos}; \citetalias{2022florian}). \citetalias{2022thomas} produced similar internal rotation velocity maps for the LMC. They observed that the rotation velocity in the southeastern region of the LMC was lower than the average and concluded that this might be attributed to either sources stripped from the SMC, or the impact of tidal pulling by the SMC. However, this phenomenon is not evident in our velocity map. They employed a machine learning algorithm to filter out foreground MW sources, allowing their final proper motion sample to include even metal-poor stars from the CMD regions (C, D, H), where MW stars would otherwise dominate. To test whether the absence of this feature in our data is indeed due to the exclusion of metal-poor stars, we cross-matched our dataset with the Gaia LMC sample and selected probable LMC stars in the C, D, and H regions. We found that our dataset contains very few sources in these regions due to the way we filter out contaminants. Therefore, the observed feature could either be a real physical effect or a result of an excess of MW stars in their sample. However, we see a small residual motion of stars towards the SMC, in the south-east region of the residual proper motion map (see Figure \ref{fig:vel_arrow}, right panel at x = --4 deg, y = --4 deg). 

\subsubsection{Residual proper motion}
\label{sec:residualPM}

The arrow plot for the residual proper motion in the sky plane is shown in the right-hand panel of Figure \ref{fig:vel_arrow}. The residual proper motion was determined by subtracting the model proper motion, derived in Section \ref{sec:velocity_model}, from the observed proper motion values. The model proper motion values were derived employing the dynamical parameters from the MCMC sampling (see Table \ref{tab:param}). The residual proper motion values were divided into two sets: one for the inner region and another for the outer region. This division was necessary due to the significant variation in the number density of sources as one moves towards the edges. The inner region was defined in the zenithal projected sky plane as spanning from $-$3 deg to 2.5 deg in the X direction and from --3 deg to 3 deg in the Y direction, while the remaining area constitutes the outer region. Each dataset was binned separately: a coarse grid covering 1.2 deg$^2$ per grid element for the outer regions, and a finer grid covering 0.14 deg$^2$ per grid element for the inner regions was utilised. This approach ensured that the number of sources used to calculate the median values in the outer region was sufficient. The average stellar density per bin was $\approx$ 24000 and $\approx$ 63700 for the outer and inner regions, respectively.

The residual map displays clear deviations from the proper motion model at various locations across the LMC disc. In our model, we assumed that the streaming motion of stars in the LMC disc is circular. However, as discussed in the previous section, stars in the bar region tend to follow elongated orbits, leading to high residuals observed around the galactic bar. \citet{2022choi} produced a similar proper motion residual map using red clump stars in the LMC, from the Gaia eDR3 dataset. They covered a much wider area, extending 10 deg from the centre of the LMC. Within the inner 6 deg, they observed a clear asymmetry in the residuals between the northern and southern regions of the disc. This feature is not apparent in our map, even when restricting our analysis to red clump stars from our sample, which is three times larger than that of \citet{2022choi}. One possible explanation for this discrepancy could be a difference in the adopted inclination angles: \citet{2022choi} derived an inclination of 23 deg, whereas our analysis yields a value of $\sim$ 33 deg. Moreover, this asymmetry is not evident in \citet{2023arranz} velocity maps and their inclination angle is comparable to our value. Another prominent feature in the residual map is the streaming motion of stars towards the bar in the LMC disc, just north of the bar. Recent N-body simulations conducted by \citet{2025rathore}, who investigated the impact of the recent LMC–SMC interaction on the LMC bar, suggest that the bar is in the process of re-aligning with the centre of the dark matter halo. Notably, the high residual velocities we observe towards the north of the bar are consistent with the resettling motions identified in their simulations, providing further support for this dynamical interpretation. In the northeastern region, the residuals clearly show motion away from the LMC centre towards the east, which could suggest a complex interaction between the LMC, the MW, and the SMC (see Section \ref{sec:Dynamical_models}). The residual map for the older population, similar to the velocity curves, is identical to the entire sample (see Figure \ref{fig:vel_arrow_old}). The high residuals in the vicinity of the bar are more pronounced in the young population, which was confined to 4 deg from the LMC centre (see Figure \ref{fig:vel_arrow_young}).

\subsubsection{Velocity maps in polar coordinates}

\begin{figure*}
    \centering
    \includegraphics[height=8cm,width=\columnwidth]{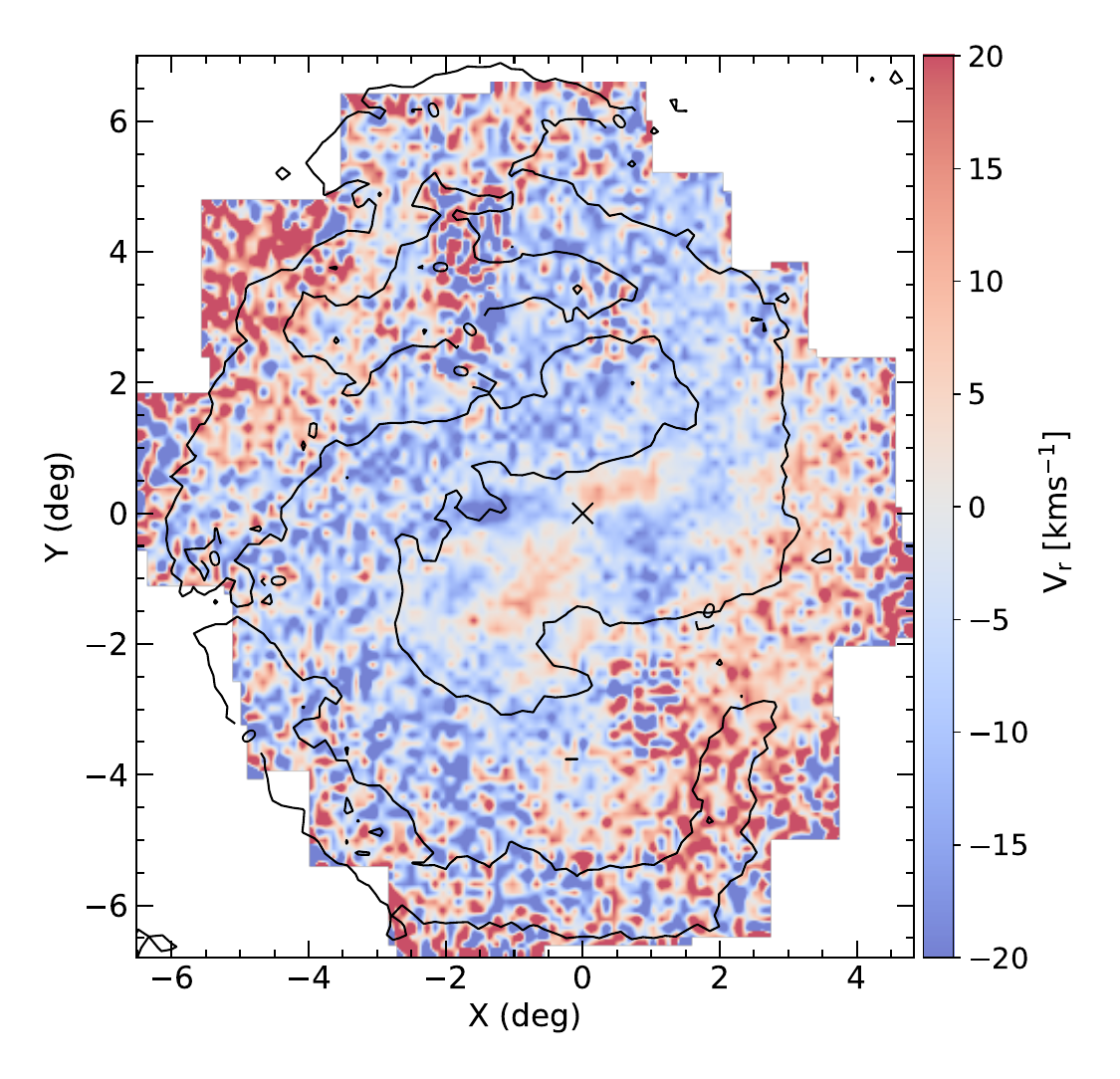}
    \hfill
    \includegraphics[height=8cm,width=\columnwidth]{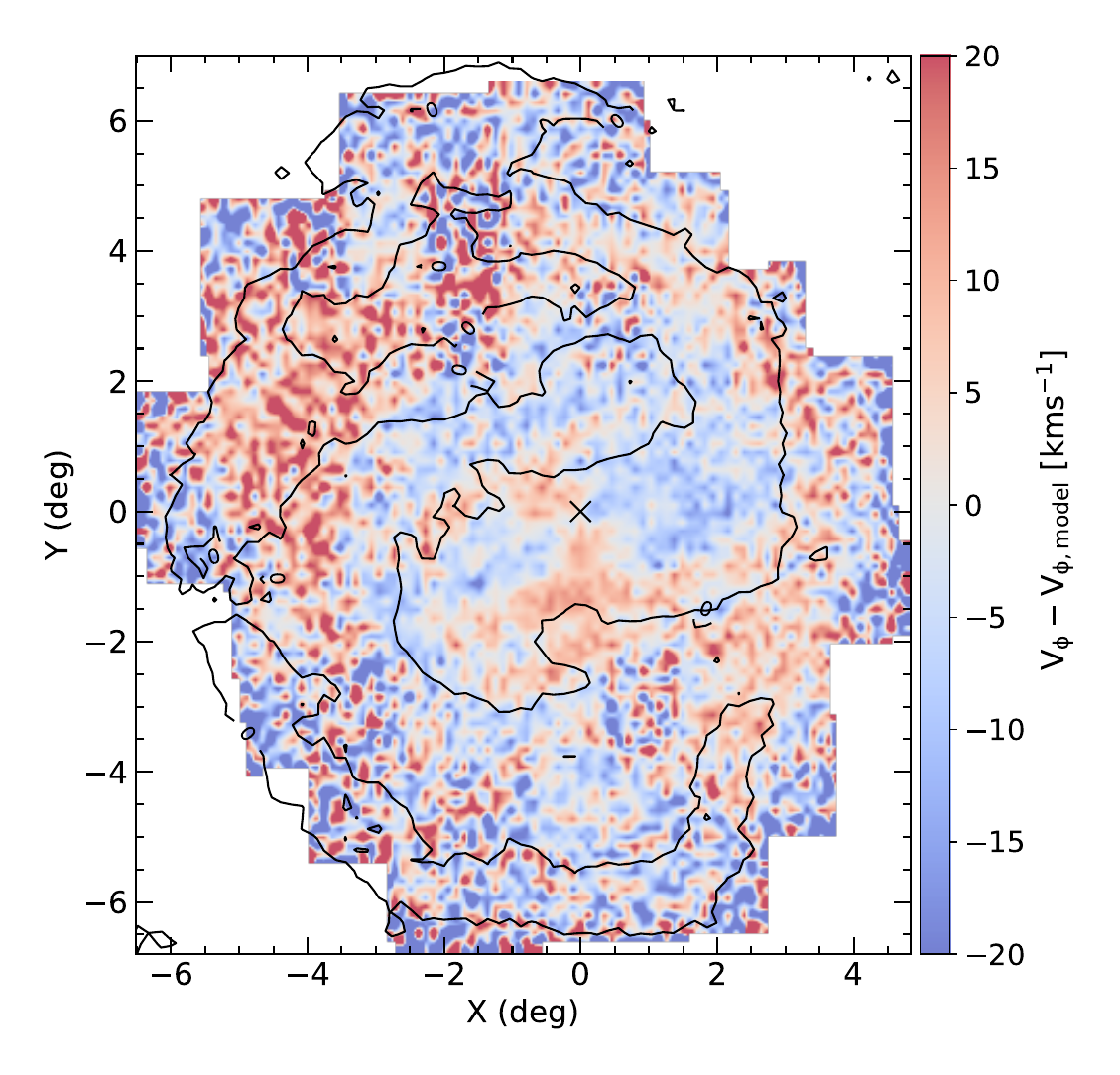}
    \caption{Internal velocity maps in the polar coordinates in the LMC disc plane aligned with the sky plane. Left: The radial velocity map.  Right: The residual azimuth velocity map (measured velocity - model velocity). The black contour line highlights the over-density regions, i.e. the bar and spiral arm. The dynamical centre derived in this study is indicated by a black cross. East is to the left and north is to the top.}
    \label{fig:vel_polar}
\end{figure*} 

The internal velocity maps in the radial and azimuth coordinates projected in the sky plane are presented in Figure \ref{fig:vel_polar}. When the coordinate system is rotated from the LMC plane to the sky plane, the polar velocities also rotate due to rotational symmetry. The radial and residual tangential velocities were discretised into 100$\times$100 bins, with an area coverage of $\approx$ 43 arcmin$^2$ per bin. The residual tangential velocity was obtained by subtracting the model tangential velocity from the observed values. The most prominent feature in both velocity maps is the quadrupole pattern observed in the bar region. This pattern arises from the asymmetric gravitational potential within the bar, attributed to the elongated stellar orbits \citep{1993sellwood}. In the residual tangential velocity map, the quadrupole pattern aligns with the major axis of the galactic bar. Similar patterns have been identified in previous studies utilising the Gaia dataset (\citetalias{2021GaiaMC}; \citealt{2023arranz}). 

In the radial velocity map, there is a noticeable increase in motion away from the LMC centre in the north-east and south-west regions, with the effect being more pronounced in the south-west. The discrepancy may be influenced by the interaction between the LMC and SMC. In \citet{2002olsen}, they studied the shape of the LMC disc utilising red clump stars and discovered an inner warp in the LMC between 2 and 4 deg from the LMC centre. This inner warp was later confirmed by \citet{2018choi} in their study of the 3D structure of the LMC, also utilising red clump stars from the Survey of the Magellanic Stellar History \citep[SMASH;][]{2017nidever}. \citet{2002olsen} found that the warp is directed towards the observer, creating the appearance of having a negative tilt compared to the southern disc. This location coincides with the south-west outward radial motion of stars in our velocity map, and hence we suggest that this outward motion could be associated with the presence of the inner warp. The deviation is more pronounced in the in-plane radial direction compared to the azimuthal direction, as the internal rotation of stars follows a clockwise direction, particularly when the motion includes a significant line-of-sight velocity component. This type of result underscores the importance of having line-of-sight velocity information for proper interpretation, which our study lacks.

The residual tangential map reveals a similar phenomenon, with eastward motion clearly visible in the north-east of the residual plot. This pattern is also evident in the residual proper motion map (see Figure \ref{fig:vel_arrow}). We attribute this high residual velocity in the azimuthal direction to a substructure located in the northeastern outskirts of the LMC, approximately 15 deg from the LMC centre, known as Eastern Substructure 1. This substructure was first reported by \citet{1955vaucouleurs} and later confirmed by \citetalias{2021GaiaMC} and \citet{2021dalal}, who identified it at about 15 deg East in the RA direction and 1 deg North in Dec direction from the LMC centre in the projected sky plane. It extends about 10 deg across in the eastward direction and is $\sim$7 deg wide in the plane of the LMC. In \citet{2022cullinane}, the kinematics of the LMC's periphery, starting at 8 deg from the centre, were studied using spectroscopic observations of red clump and RGB stars from the Magellanic Edges Survey \citep[MagES;][]{2020cullinane}. They reported that the inner region of the northeastern substructure is dynamically stable, with its tracers located in the same plane as the LMC disc (scale height, z $\approx$ 0). A similar study in this region by \citet{2023navarrete}, which included a more extended area and utilised Mira variable stars as tracers, concluded that this substructure results from a perturbed disc. Both studies simulated possible interaction histories between the LMC, SMC, and MW, suggesting at least three crossings of the SMC. In \citet{2023navarrete}, their simulation indicated an outward radial in-plane motion of stars in the northeastern region. Hence, we suggest that this excess tangential motion towards the north-east of the LMC could be a kinematical indicator of this substructure from within the LMC disc plane. 

\begin{figure}
    \centering
    \includegraphics[height=8.5cm,width=\columnwidth]{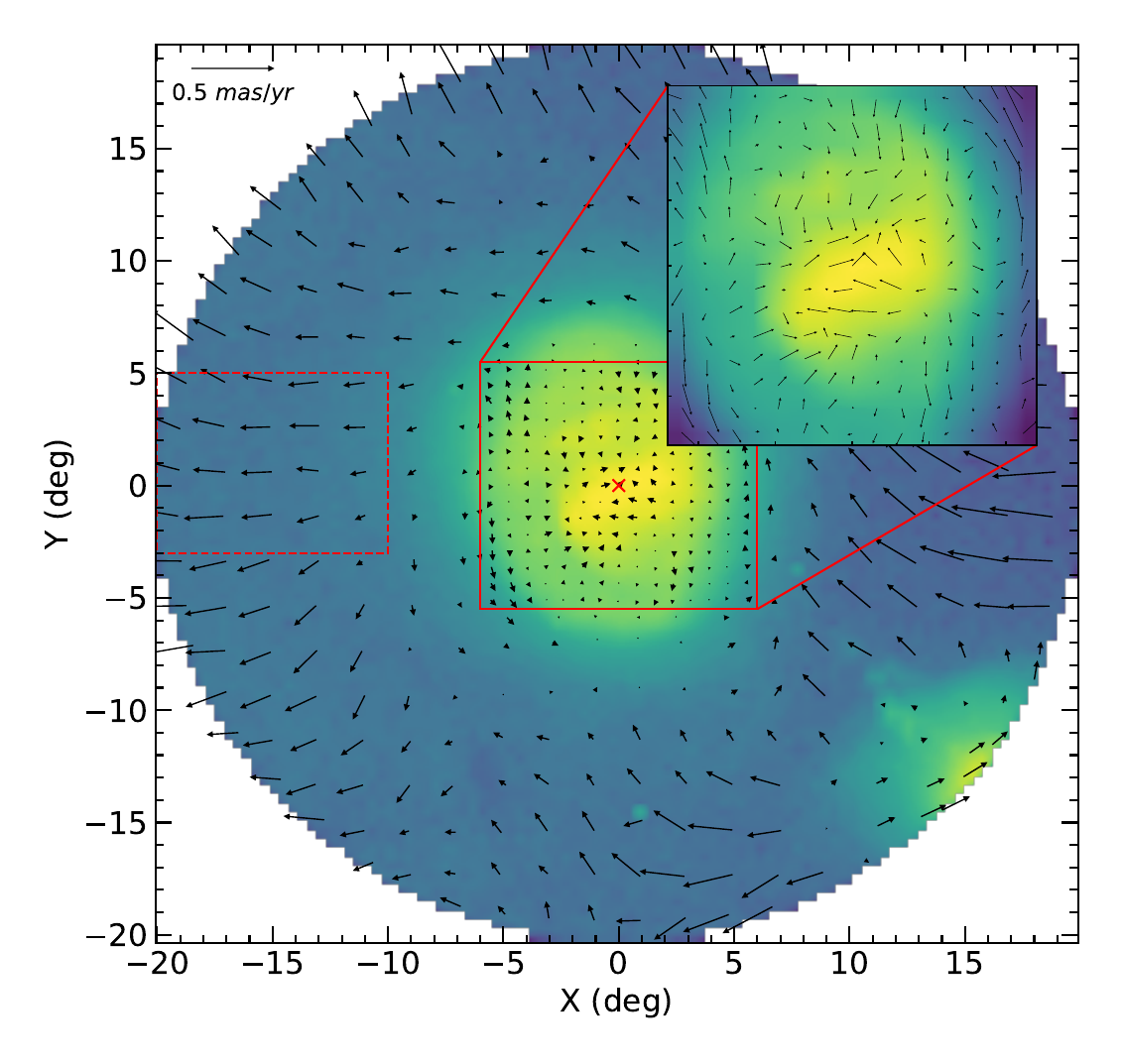}
    \caption{Vector diagram of the residual proper motion for probable LMC sources in the Gaia eDR3 dataset. Due to the applied selection criteria \citepalias{2021GaiaMC}, portions of the SMC and the Magellanic Bridge region are also visible. The dynamical centre derived by \citetalias{2021GaiaMC} is marked with a red cross. The inset shows the zoomed-in view of the inner 6$^\degree$, and the dashed rectangular box indicates the location of the Eastern Substructure 1.} 
    \label{fig:vel_gaia}
\end{figure}

To explore this connection in detail, we constructed a residual velocity map similar to Figure \ref{fig:vel_arrow} for probable LMC sources in Gaia eDR3 data (see Section \ref{sec:CFR}), as shown in Figure \ref{fig:vel_gaia}. The VMC survey covers up to a radius of 6 deg around the LMC centre, which restricts our ability to visually correlate the residual motion with Eastern substructure 1 using VMC data alone. In contrast, Gaia offers a broader coverage radius of 20 deg around the LMC centre. The resulting residual map shows the eastward deflection of stars within the inner region, which extends towards the location of the substructure as marked in Figure \ref{fig:vel_gaia} (inset and red box indicate the inner region and the substructure, respectively). This provides the first observational evidence of the kinematic connection between the residual motion and the substructure.

\section{Comparison with dynamical models}
\label{sec:Dynamical_models}

\subsection{Model description}
\label{sec:Dynamical_models_intro}

To further investigate the nature of the residual eastward motion, we employed a suite of simple dynamical models which simulate the interaction of the LMC with the MW and the SMC. The models are those presented by \citet{2022cullinaneA} and were developed to study the substructures in the LMC periphery. In brief, the LMC is modelled as a system of particles within a rigid two-component potential, representing the disc and the halo. The disc is modelled as an exponential potential with a disc mass of 2 $\times$ $10^9$ $\mathrm{M}_{\odot}$, a scale radius of 1.5 kpc and a scale height of 0.4 kpc. The dark matter halo follows a Hernquist profile \citep{1990hernquist}, with a mass of 1.5 $\times$ $10^{11}$ $\mathrm{M}_{\odot}$ and scale radius of 20 kpc, giving a circular velocity of $\sim$90 km s$^{-1}$ at 10 kpc. The MW is represented by the three-component potential \texttt{MWPotential2014} from \citet{2015bovy}, while the SMC is represented as a Hernquist profile with a mass of 2.5 $\times$ $10^9$ $\mathrm{M}_{\odot}$ and a scale radius of 0.043 kpc, having a circular velocity of 60 km s$^{-1}$ at 2.9 kpc. To establish initial conditions, the orbits of the LMC, SMC, and MW were rewound from their present-day positions back to 1 Gyr, ensuring a physically motivated setup. The orientation of the LMC disc was set based on the values from \citet{2018choi} and remains fixed throughout the simulation due to the rigid potential. The LMC disc was initialised with approximately 2.5 $\times$ $10^6$ tracer particles using the AGAMA software package \citep{2019vasiliev}, whereas no tracer particles were assigned to the SMC. The system is then evolved forward from 1 Gyr ago to the present day. Since the models were originally designed to study the LMC periphery, they include only particles whose apocentres were greater than 7 kpc to improve the computational efficiency.

A set of five different model suites was computed, each representing a different galaxy mass configuration. For this study, we selected two model suites: their `base-case' model, which uses common literature values for the masses of the LMC, SMC, and MW, and a variant of the same model excluding the SMC. The base-case model was run for 100 realisations of LMC-MW-SMC interactions, sampling the measured present-day systemic velocity and line-of-sight distances of the LMC and SMC within Gaussian uncertainties (see Table 7 of \citealt{2022cullinaneA}). This approach generates an ensemble of possible orbital histories for the Clouds. The majority of model realisations (87 instances) consistently show a recent close pericentric passage of the SMC around the LMC $\sim$150 Myr ago\footnote{We used a pericentre passage time threshold of less than 200 Myr; the remaining model realisations have the most recent SMC pericentric passage occurring at older times.} at a distance of $\sim$7.7 kpc, occurring below the LMC disc plane at z\footnote{z represents the out-of-plane distance, with a negative z indicating a position behind the LMC disc plane relative to the observer.} $\sim -$6.2 kpc, with the projected pericentre at $\sim$4.5 kpc in the south-west direction. Additionally, around 30 model realisations show a second SMC disc crossing $\sim$400 Myr ago at a broad range of in-plane distances ($\sim$21 kpc). A third crossing,  occurring $\sim$900 Myr ago, is present in four model instances, though this fraction would increase if the simulation were rewound longer. Additionally, two realisations exhibit a second pericentric passage $\sim$900 Myr ago at 6.9 kpc, with a smaller out-of-plane distance (z $\sim$ 2.3 kpc); these statistics would also increase with a longer rewind time.

\subsection{Residual eastward motion}

\begin{figure*}
    \centering
    \includegraphics[height=8cm,width=\columnwidth]{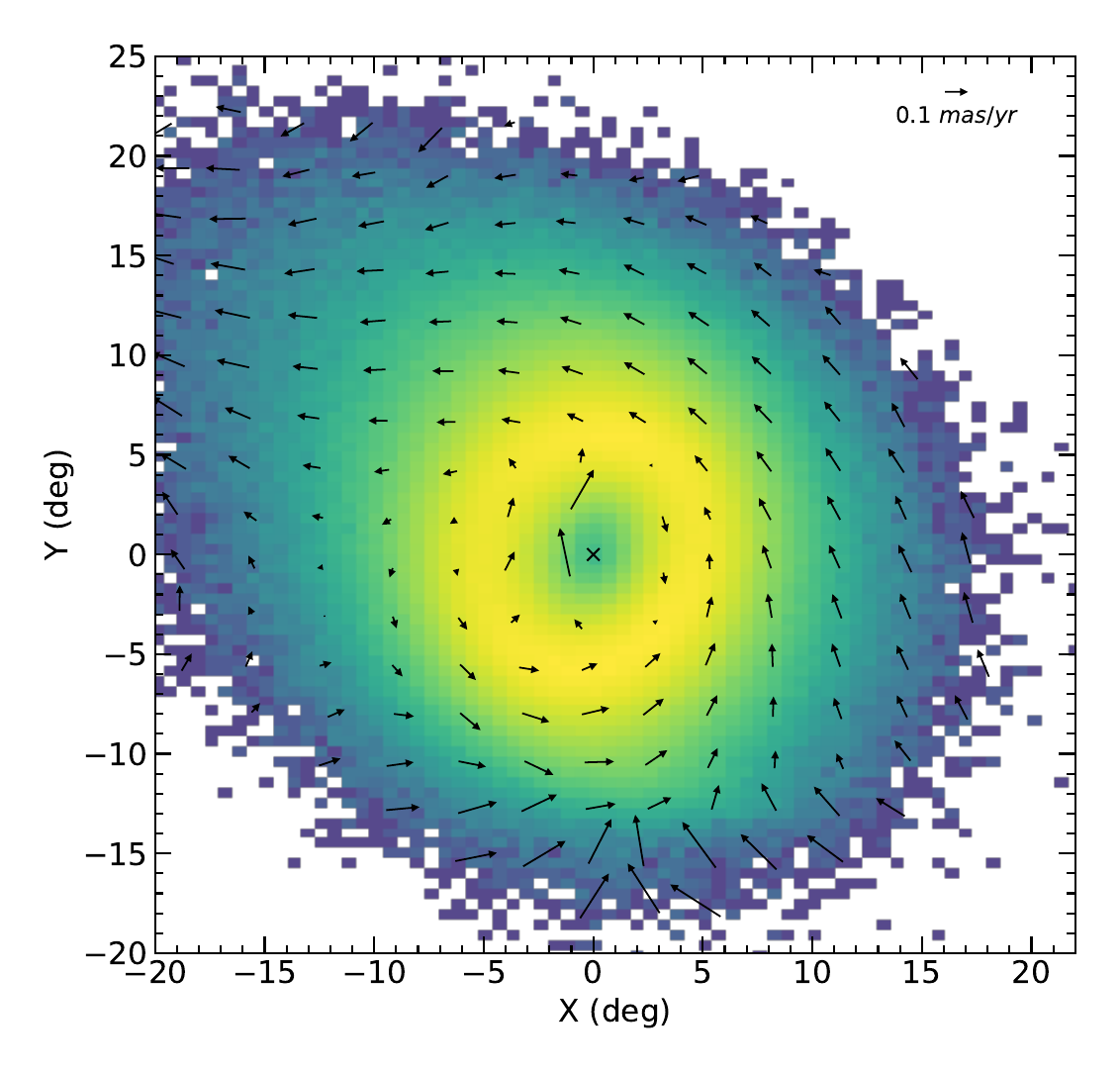}
    \hfill
    \includegraphics[height=8cm,width=\columnwidth]{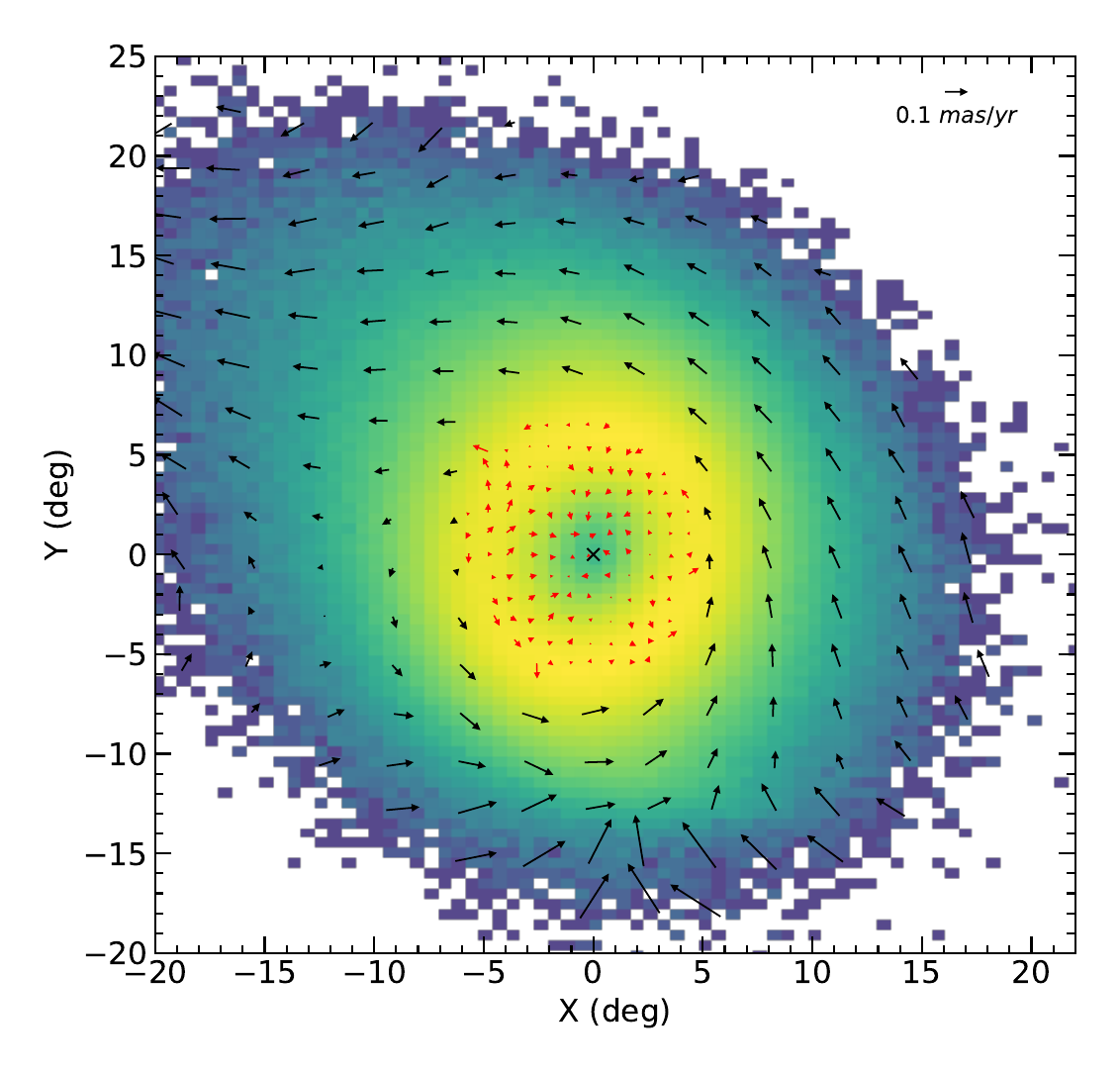}
\caption{Vector diagram of residual proper motion in the LMC disc plane, aligned with the sky plane, for a single realisation of the base case dynamical model. Left: Residual proper motion for one realisation exhibiting the motion towards the east. The reduced number density in the inner region results from considering only sources with an apocentre greater than 7 kpc. Right: The same residual map, with the inner 5 deg overlaid with residual motion from VMC data shown in red arrows. A reference length for the vectors is shown. The background displays the density of particles per bin in a 100$\times$100 grid, from the simulation. The dynamical centre, taken from \citet{2022cullinane}, is marked by a black cross. East is to the left, and north is to the top.}
    \label{fig:vel_arrow_dyn}
\end{figure*} 

The simulations were run in a Cartesian coordinate system where the origin is the location of the MW at present and mock observations were generated on the final snapshot, with the LMC in its present-day location, in the frame of reference of the Sun. These mock observations include information on the 3D phase space, as distances to individual particles can be directly estimated from the simulation, unlike in actual observations. As done for the VMC observations (see section \ref{sec:int_kinematics}), a de-projection from the sky plane to the LMC disc plane was performed on the mock data. The residual proper motion was derived using the same procedure as detailed in \ref{sec:residualPM}. However, the rotation curve was produced using the rigid potential of the LMC described in section \ref{sec:Dynamical_models_intro}, since the simulation was run with the same potential.

Subsequently, we created vector diagrams of the residual velocity maps for the mock observations, extending from the vicinity of that seen in the VMC data, to the region of the Eastern Substructure 1, and identified realisations that exhibit the eastward motion. Approximately 49 per cent of realisations in the `base-case' display some degree of residual eastward motion. Figure \ref{fig:vel_arrow_dyn} presents the residual proper motion of the mock data for a realisation that clearly exhibits eastward motion, with the VMC residual motion shown within the inner 5 deg in the right panel. About 17 per cent of the realisations exhibit no eastward motion, while the remaining 31 per cent display a slight residual motion towards the north-east. The remaining three realisations show distinct residual patterns and a perturbed disc structure. Among the simulation instances that exhibit significant residual motion to the east, 14 per cent experienced a disc crossing at $\sim$400 Myr ago, along with the most recent pericentric passage,  while 26 per cent underwent multiple SMC crossings. The majority of realisations displaying residual eastward motion have only undergone the most recent pericentric passage, typically with pericentric distances towards the larger end of the full range simulated. We also generated residual maps for the model that excludes the SMC to isolate the influence of the MW alone. Due to computational constraints, this model was limited to 12 realisations, each incorporating the full particle distribution. For realisations where the base model exhibited residual motion towards the east, the SMC-excluded model showed a comparable overall residual pattern but with increased residual amplitude. These findings suggest that the observed residual motion arises from the combined effects of the tidal influence of the MW, shaped by the LMC disc’s dynamical response to the recent SMC passage. 

An important limitation of the dynamical modelling is its use of rigid potentials, which consequently results in the exclusion of self-gravity. As discussed in \citet{2022cullinaneA}, this prevents the deformation of the galaxies’ dark matter halos -- which can affect their global orbits -- as well as limiting the response of the LMC’s stellar disc to interactions which would otherwise perturb it. Given that the residual eastward motion is observed at relatively small LMC radii (compared to the size of its dark matter halo), the deformation of the LMC’s halo in response to the MW is unlikely to significantly affect the residual motion directly; however it may affect the LMC’s orbit around the MW and thus the amplitude of the observed residual motion.

More significant is the effect of these limitations on interactions between the Clouds. Admittedly, the specific interactions captured in these model suites are not those which are likely to cause the strongest effects on the LMC disc -- namely, the $\sim$400 Myr crossing of the LMC’s disc by the SMC occurs at a large LMC galactocentric radius, and the recent SMC pericentric passage ($\sim$150 Myr ago) occurred at a substantial out-of-plane distance. However, the SMC’s orbit is influenced by dynamical friction and dark matter wakes in the LMC’s halo not captured in the simpler models, which could make its orbit more eccentric, introducing further uncertainty in the SMC’s orbit and its interactions with the LMC.  

Finally, we reiterate that these simple models do not include particles with apocentres < 7 kpc -- which overlaps the innermost region in which we observe the residual eastern motion in the VMC data -- and we therefore caution over-interpretation of the kinematics near the model centres. It is therefore clear that the origin of the residual eastern motion cannot be definitively determined using the simplified dynamical modelling employed here, with more realistic models required to fully understand how it is linked to the complex interaction history of the LMC.

\section{Summary and conclusion}
\label{sec:conclusion}

In this study, we have assessed the internal kinematics of the LMC by calculating stellar proper motions using near-infrared data from the VMC survey. The survey takes multiple epochs in the K$_s$ band, averaging around 14 epochs per tile, with a time baseline of approximately 10 years. This represents a substantial improvement, more than tripling the time baseline compared to previous proper motion studies of the LMC using VMC data (\citetalias{2022florian}; \citetalias{2022thomas}). Consequently, we achieved high-precision proper motion measurements, reducing the uncertainty from 6 mas yr$^{-1}$ to 1.5 mas yr$^{-1}$ for LMC sources.

We derived the dynamical parameters of the LMC by fitting the measured proper motions to analytical models of the galaxy's velocity field. These parameters were calculated for the entire sample, as well as separately for the young and old populations. Most of the dynamical parameters were obtained without degeneracy using the transverse velocity measurements. The dynamical centre of the LMC derived in this study, ($\alpha_0,\delta_0 = 80.30^\circ \substack{+0.05 \\ -0.05}, -69.27^\circ \substack{+0.02 \\ -0.02}$), is closer to the centre of the {H\,\sc{i}} gas and deviates from the photometric centre, consistent with findings from Gaia and dynamical modelling of the LMC. The inclination angle ($i = 32.8^\circ \substack{+0.2 \\ -0.2}$) and the position angle of the line-of-nodes ($\Theta = 134.5^\circ \substack{+0.5 \\ -0.5}$), and the proper motion of the COM, ($\mu_{W,0}, \mu_{N,0} = -1.854 \substack{+0.001 \\ -0.001} \mathrm{mas yr}^{-1}, 0.330 \substack{+0.003 \\ -0.003} \mathrm{mas yr}^{-1}$), are consistent with values reported in the literature. 

The derived dynamical parameters were utilised to generate velocity curves and maps in both the plane of the LMC and the sky plane. We derived the velocity curves for both tangential and radial motions. The rotation curve exhibited a dark halo model trend, with a scale radius of $R_0 = 2.94$ kpc and asymptotic velocity of $V_0 = 91.94$ km s$^{-1}$. The bump in the rotation curve and its decline beyond 5.5 kpc indicate the effect of the central bar and the elliptical nature of the LMC. Analysing the shift in rotation curves between the young and old stellar populations in the LMC confirms the previous findings of an asymmetric drift. This highlights that older populations are kinematically hot, while younger populations remain dynamically cold, reflecting the progressive heating of stellar orbits over time. The radial velocity curve aligns with previous studies, revealing opposing trends between the young and old populations. In the young population, the increase in radial velocity within the bar suggests deviations from circular orbits, pointing to the bar's influence on gas dynamics.

The velocity maps encompass both the internal rotation motion and the residual motion within the sky plane. The internal rotation map clearly displays the clockwise motion of LMC sources about the dynamical centre. We observed elongated orbits for sources in the bar, a characteristic previously identified by \citetalias{2022florian}. This deviation from circular orbits is also apparent in the residual map. A significant residual motion of stars towards the east, away from the LMC centre, is evident in the residual tangential velocity map and is kinematically associated with a substructure in the LMC periphery known as Eastern Substructure 1. Comparison with simple dynamical models suggests this motion is driven by tidal forces from the MW, tempered by the recent pericentric passage of the SMC around the LMC; but further quantitative analysis of more realistic models are required to confirm the motion's origin.

To effectively study the internal kinematics of the LMC, improved methods for removing MW sources are essential. Recent advancements have involved using machine learning algorithms to refine catalogues (\citetalias{2022thomas}; \citealt{2023arranz}). While consensus on the LMC’s dynamical parameters is emerging, further refinement is needed, particularly by incorporating the ellipticity and the non-axisymmetric nature of the LMC into velocity models. Although radial velocity data were more prevalent than proper motion measurements in the past, the 21st century has seen a lack of high-precision radial velocity measurements. Upcoming spectroscopic surveys, such as the SDSS-V Local Volume Mapper \citep[LVM;][]{2017kollmeier} and the 4MOST telescope's One Thousand and One Magellanic Fields \citep[1001MC;][]{2019cioni} survey, will bridge existing gaps by providing extensive radial velocity data for the Magellanic Clouds. Finally, to explore the interaction history between the Magellanic Clouds and the MW, dynamical modelling and numerical simulations \citep{2019erkal, 2019tepper, 2019garavito, 2020lucchini, 2021lucchini, 2021petersen, 2023vasiliev, 2024aarranz} that incorporate observational parameters are crucial for interpreting the observational data within a theoretical framework.

\section*{Data availability}

The proper motion tables for all VMC tiles will be included in the final data release of the VMC survey (DR7), which will be available through the VMC ESO Phase3 collection.\footnote{\url{https://doi.eso.org/10.18727/archive64/}} For further details regarding the data release, see Section 4.3 of \citet{2025cioni}.

\begin{acknowledgements}
We thank the Cambridge Astronomy Survey Unit (CASU) and the
Wide Field Astronomy Unit (WFAU) in Edinburgh for providing
calibrated data products under the support of the Science and
Technology Facility Council (STFC). This research was funded by the Deutsche Forschungsgemeinschaft (DFG, German Research Foundation) - Cl 213/10-1.
FN acknowledges funding by DLR grant 50 OR 2216.  
This study is based on observations obtained
with VISTA at the Paranal Observatory under programmes 179.B-2003, 0100.C-0248, 106.2107, 108.222A, 109.231H, 110.259F. For the purpose of open access, the author has applied a Creative Commons Attribution (CC BY) licence to any Author Accepted Manuscript version arising from this submission. This work has made use of data from the European Space Agency (ESA)
mission Gaia (\url{https://www.cosmos.esa.int/gaia}), processed by the Gaia
Data Processing and Analysis Consortium (DPAC, \url{https://www.cosmos.esa.int/web/gaia/dpac/consortium}). Funding for the DPAC has been provided by national institutions, in particular the institutions participating in the Gaia Multilateral Agreement. This work made extensive use of several Python libraries like Astropy, a community-developed core Python package for Astronomy \citep{astropy:2013,astropy:2018, astropy:2022}, Matplotlib \citep{matplotlib}, SciPy \citep{scipy} and NumPy \citep{numpy}.
\end{acknowledgements}

\bibliographystyle{aa}
\bibliography{biblio.bib}

\onecolumn
\begin{appendix}

\section{Overview of VMC tiles}

\begin{table*}[h!]\scriptsize
\centering
\caption{Details of all 68 LMC tiles in the VMC survey.}
\begin{tabular}{cccrccc}
\hline\hline
\noalign{\smallskip}
Tile & RA$_{\mathrm{J}2000}$ & Dec$_{\mathrm{J}2000}$ & Position Angle & Number of Epochs & Old Time baseline & New Time baseline \\ 
 & (hh:mm:ss) & (deg:mm:ss) & (deg) & & (years) & (years) \\
\noalign{\smallskip}
\hline
\noalign{\smallskip}
LMC 2\_3 & 04:48:04.752 & $-$74:54:11.880 &  $-$101.226 & 14 & 4.25 & 9.33\\
LMC 2\_4 & 05:04:42.696 & $-$75:04:45.120 &  $-$97.3198 & 14 & 3.16 & 9.33\\
LMC 2\_5 & 05:21:38.664 & $-$75:10:50.160 &  $-$93.3381 & 14 & 4.25 & 9.33\\
LMC 2\_6 & 05:38:43.056 & $-$75:12:21.240 &  $-$89.3214 & 13 & 4.50 & 9.33\\
LMC 2\_7 & 05:55:45.720 & $-$75:09:17.280 &  $-$85.3118 & 14 & 2.25 & 9.25\\
LMC 3\_2 & 04:37:05.256 & $-$73:14:30.120 & $-$103.7726 & 14 & 4.25 & 9.33\\
LMC 3\_3 & 04:51:59.640 & $-$73:28:09.120 & $-$100.284 & 14 & 3.33 & 9.25\\
LMC 3\_4 & 05:07:14.472 & $-$73:37:49.800 &  $-$96.7101 & 14 & 5.00 & 9.25\\
LMC 3\_5 & 05:22:43.056 & $-$73:43:25.320 &  $-$93.0788 & 14 & 1.92 & 12.16\\
LMC 3\_6 & 05:38:18.096 & $-$73:44:51.000 &  $-$89.4206 & 14 & 4.42 & 9.25\\
LMC 3\_7 & 05:53:51.912 & $-$73:42:05.760 &  $-$85.7675 & 14 & 3.33 & 9.25\\
LMC 3\_8 & 06:09:16.920 & $-$73:35:12.120 &  $-$82.1511 & 14 & 4.25 & 9.16\\
LMC 4\_2 & 04:41:30.768 & $-$71:49:16.320 & $-$102.7172 & 11 & 3.08 & 13.16\\
LMC 4\_3 & 04:55:19.512 & $-$72:01:53.400 &  $-$99.4885 & 12 & 3.08 & 13.16\\
LMC 4\_4 & 05:09:32.496 & $-$72:10:16.680 &  $-$96.2767 & 14 & 3.00 & 9.00\\
LMC 4\_5 & 05:23:46.560 & $-$72:15:21.960 &  $-$92.4781 & 14 & 2.08 & 8.42\\
LMC 4\_6 & 05:38:00.408 & $-$72:17:20.040 &  $-$89.4906 & 14 & 4.08 & 12.25\\
LMC 4\_7 & 05:50:50.496 & $-$72:15:39.960 &  $-$86.8141 & 14 & 2.08 & 8.33\\
LMC 4\_8 & 06:03:40.872 & $-$72:10:06:240 &  $-$83.4749 & 14 & 2.83 & 10.00\\
LMC 4\_9 & 06:17:43.560 & $-$72:00:48.240 &  $-$80.1877 & 14 & 4.16 & 9.00\\
LMC 5\_1 & 04:31:28.032 & $-$70:06:57.600 & $-$105.0473 & 14 & 2.83 & 9.16\\
LMC 5\_2 & 04:44:01.728 & $-$70:22:21.000 & $-$102.1178 & 14 & 2.92 & 8.16\\
LMC 5\_3 & 04:56:52.488 & $-$70:34:25.680 &  $-$99.1173 & 14 & 3.08 & 10.00\\
LMC 5\_4 & 05:10:41.543 & $-$70:43:05.880 &  $-$96.0612 & 14 & 2.92 & 8.00\\
LMC 5\_5 & 05:24:30.336 & $-$70:48:34.200 &  $-$92.6525 & 14 & 1.33 & 11.08\\
LMC 5\_6 & 05:36:53.928 & $-$70:49:52.320 &  $-$89.8559 & 14 & 4.00 & 8.00\\
LMC 5\_7 & 05:49:43.944 & $-$70:47:54.960 &  $-$86.7456 & 14 & 2.08 & 8.08\\
LMC 5\_8 & 06:02:56.232 & $-$70:42:25.920 &  $-$83.655  & 14 & 4.42 & 9.25\\
LMC 5\_9 & 06:15:59.112 & $-$70:33:27.360 &  $-$80.6038 & 14 & 4.58 & 9.33\\
LMC 6\_1 & 04:36:49.488 & $-$68:43:50.880 & $-$103.7944 & 14 & 2.92 & 9.08\\
LMC 6\_2 & 04:48:39.072 & $-$68:57:56.520 & $-$101.0355 & 14 & 2.92 & 9.00\\
LMC 6\_3 & 05:00:42.216 & $-$69:08:54.240 &  $-$98.2198 & 14 & 2.75 & 8.92\\
LMC 6\_4 & 05:12:55.800 & $-$69:16:39.360 &  $-$95.3605 & 14 & 1.16 & 12.25\\
LMC 6\_5 & 05:25:16.272 & $-$69:21:08.280 &  $-$92.4724 & 14 & 2.92 & 9.00\\
LMC 6\_6 & 05:37:40.008 & $-$69:22:18.120 &  $-$89.5708 & 14 & 1.26 & 11.42\\
LMC 6\_7 & 05:50:03.168 & $-$69:20:09.240 &  $-$86.6715 & 14 & 2.00 & 6.42\\
LMC 6\_8 & 06:02:21.984 & $-$69:14:42.360 &  $-$83.7904 & 14 & 3.25 & 12.00\\
LMC 6\_9 & 06:14:32.832 & $-$69:05:59.640 &  $-$80.9426 & 13 & 4.75 & 9.00\\
LMC 6\_10 & 06:26:32.280 & $-$68:54:05.760 & $-$78.1423 & 14 & 4.16 & 9.00\\
LMC 7\_1 & 04:40:09.167 & $-$67:18:19.800 & $-$103.0233 & 14 & 2.92 & 9.00\\
LMC 7\_2 & 04:51:17.832 & $-$67:31:39.000 & $-$100.4214 & 14 & 2.08 & 8.25\\
LMC 7\_3 & 05:02:55.200 & $-$67:42:14.760 &  $-$97.7044 & 14 & 2.25 & 12.08\\
LMC 7\_4 & 05:14:06.384 & $-$67:49:21.720 &  $-$95.0871 & 14 & 2.08 & 8.25\\
LMC 7\_5 & 05:25:58.440 & $-$67:53:42.000 &  $-$92.3088 & 18 & 7.00 & 7.00\\
LMC 7\_6 & 05:37:17.832 & $-$67:54:47.880 &  $-$89.6572 & 14 & 2.92 & 8.92\\
LMC 7\_7 & 05:48:54.000 & $-$67:52:51.240 &  $-$86.9403 & 14 & 1.16 & 10.16\\
LMC 7\_8 & 06:00:27.696 & $-$67:47:48.120 &  $-$84.2339 & 14 & 4.16 & 9.00\\
LMC 7\_9 & 06:11:54.384 & $-$67:39:41.400 &  $-$81.5567 & 13 & 4.75 & 8.25\\
LMC 7\_10 & 06:23:11.736 & $-$67:28:34.320 & $-$78.9185 & 14 & 4.16 & 8.25\\
LMC 8\_2 & 04:54:11.568 & $-$66:05:47.760 &  $-$99.7547 & 14 & 3.00 & 9.08\\
LMC 8\_3 & 05:04:53.952 & $-$66:15:29.880 &  $-$97.2489 & 12 & 2.08 & 13.25\\
LMC 8\_4 & 05:15:43.464 & $-$66:22:19.920 &  $-$94.7132 & 14 & 4.08 & 10.08\\
LMC 8\_5 & 05:26:37.152 & $-$66:26:15.720 &  $-$92.1598 & 14 & 2.00 & 9.08\\
LMC 8\_6 & 05:37:34.104 & $-$66:27:15.840 &  $-$89.5932 & 14 & 2.16 & 8.83\\
LMC 8\_7 & 05:48:30.120 & $-$66:25:19.920 &  $-$87.0304 & 14 & 4.25 & 8.83\\
LMC 8\_8 & 05:59:23.136 & $-$66:20:28.680 &  $-$84.4802 & 15 & 1.08 & 13.16\\
LMC 8\_9 & 06:10:10.632 & $-$66:12:43.560 &  $-$81.9529 & 14 & 4.25 & 9.08\\
LMC 9\_3 & 05:06:40.632 & $-$64:48:40.320 &  $-$96.8439 & 12 & 3.83 & 13.25\\
LMC 9\_4 & 05:16:39.792 & $-$64:54:59.760 &  $-$94.5007 & 14 & 3.66 & 8.92\\
LMC 9\_5 & 05:26:58.512 & $-$64:58:45.840 &  $-$92.0799 & 14 & 3.75 & 9.08\\
LMC 9\_6 & 05:37:19.104 & $-$64:59:45.240 &  $-$89.6513 & 14 & 4.50 & 9.00\\
LMC 9\_7 & 05:47:55.128 & $-$64:57:52.920 &  $-$87.1624 & 14 & 2.92 & 10.75\\
LMC 9\_8 & 05:57:57.168 & $-$64:53:24.360 &  $-$84.8071 & 14 & 4.66 & 9.66\\
LMC 9\_9 & 06:08:10.343 & $-$64:46:05.880 &  $-$82.4095 & 14 & 4.08 & 9.66\\
LMC 10\_4 & 05:17:46.656 & $-$63:27:46.440 & $-$94.249 & 14 & 4.66 & 8.16\\
LMC 10\_5 & 05:27:33.096 & $-$63:31:19.200 & $-$91.9491 & 14 & 2.75 & 9.25\\
LMC 10\_6 & 05:37:22.848 & $-$63:32:13.560 & $-$89.6358 & 14 & 4.08 & 8.83\\
LMC 10\_7 & 05:47:11.424 & $-$63:30:29.520 & $-$87.3272 & 14 & 3.16 & 8.16\\
\noalign{\smallskip}
\hline
\label{tab:Tiles}
\end{tabular}
\tablefoot{The RA and Dec columns give the central coordinates of the tiles.}
\end{table*} 

\newpage
\section{Internal kinematics of young and old stellar population}

\begin{figure*}[h!]
    \centering
    \includegraphics[width=9cm]{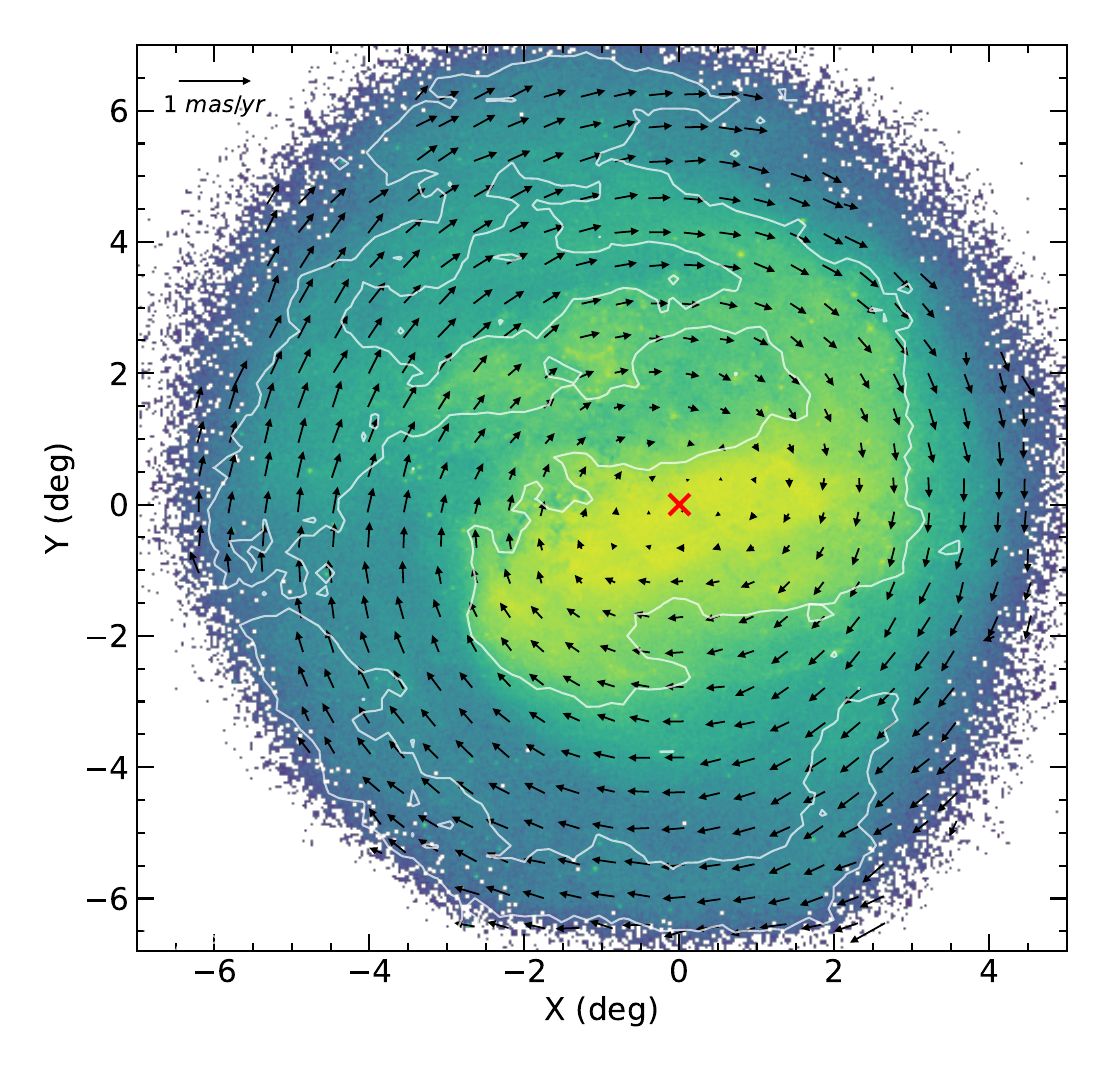}
    \hfill
    \includegraphics[width=9cm]{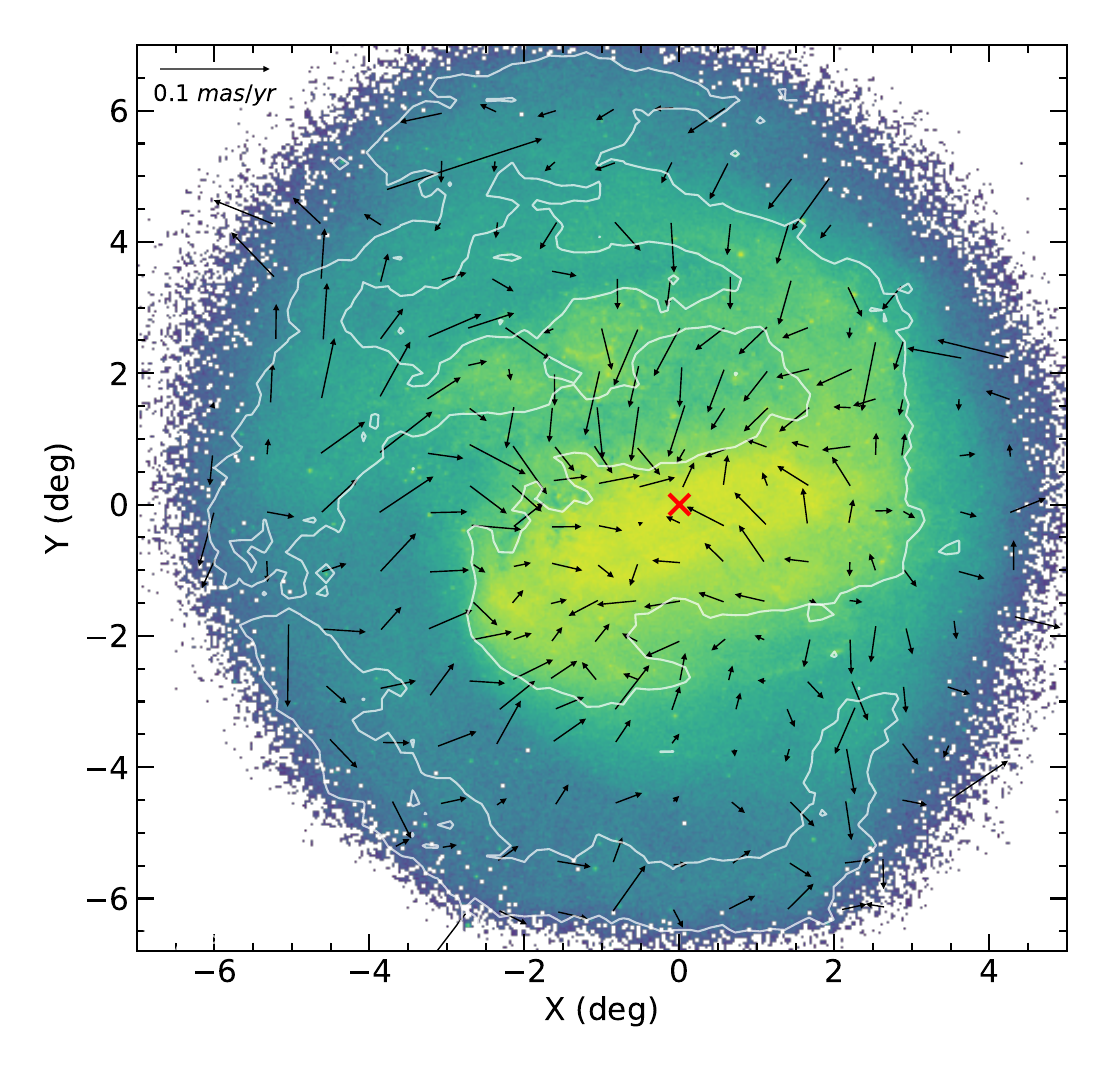}
    \caption{Same as in Figure \ref{fig:vel_arrow}, but for the old population.} 
    \label{fig:vel_arrow_old}
\end{figure*} 

\begin{figure*}[h!]
    \centering
    \includegraphics[width=9cm]{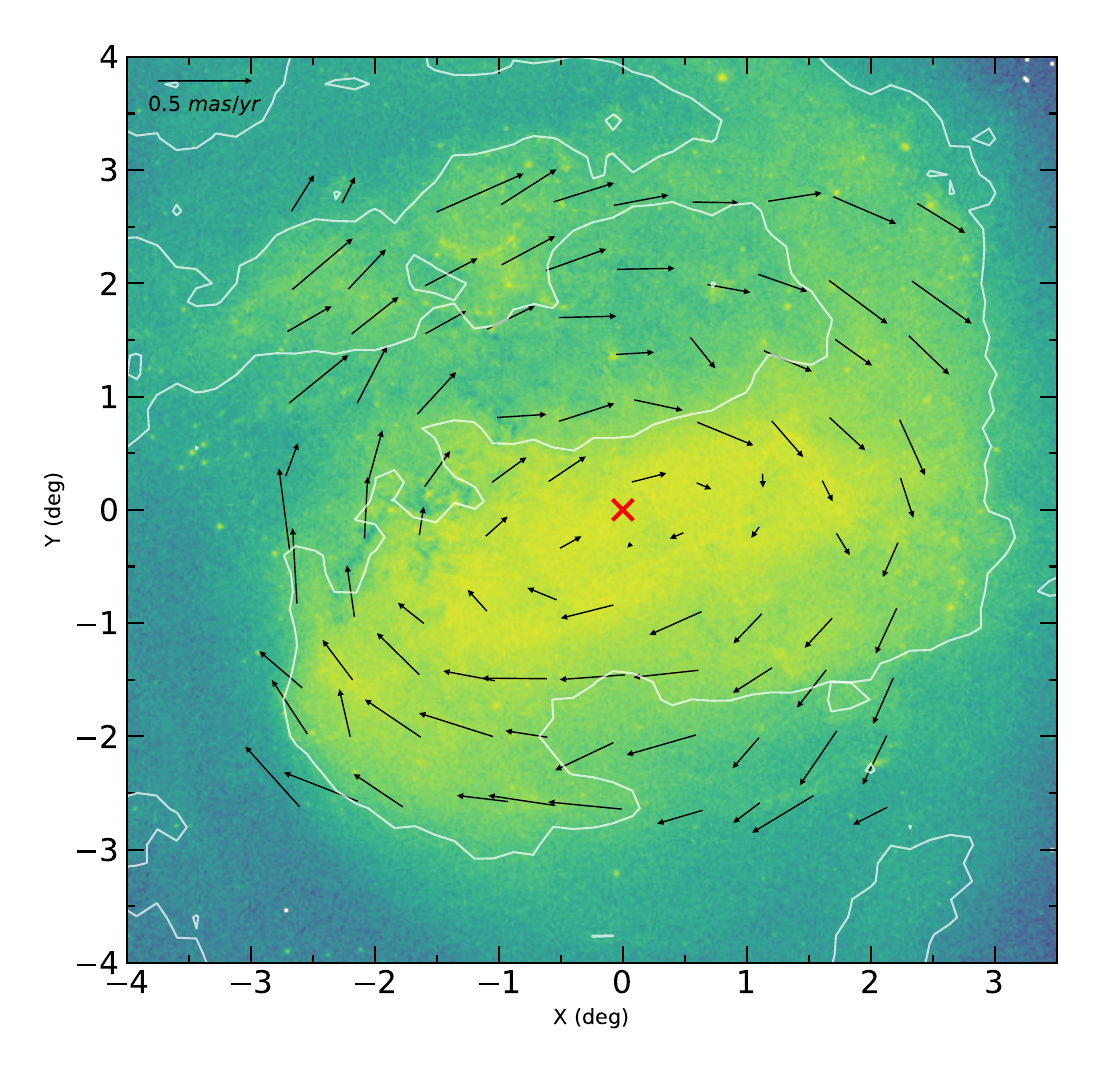}
    \hfill
    \includegraphics[width=9cm]{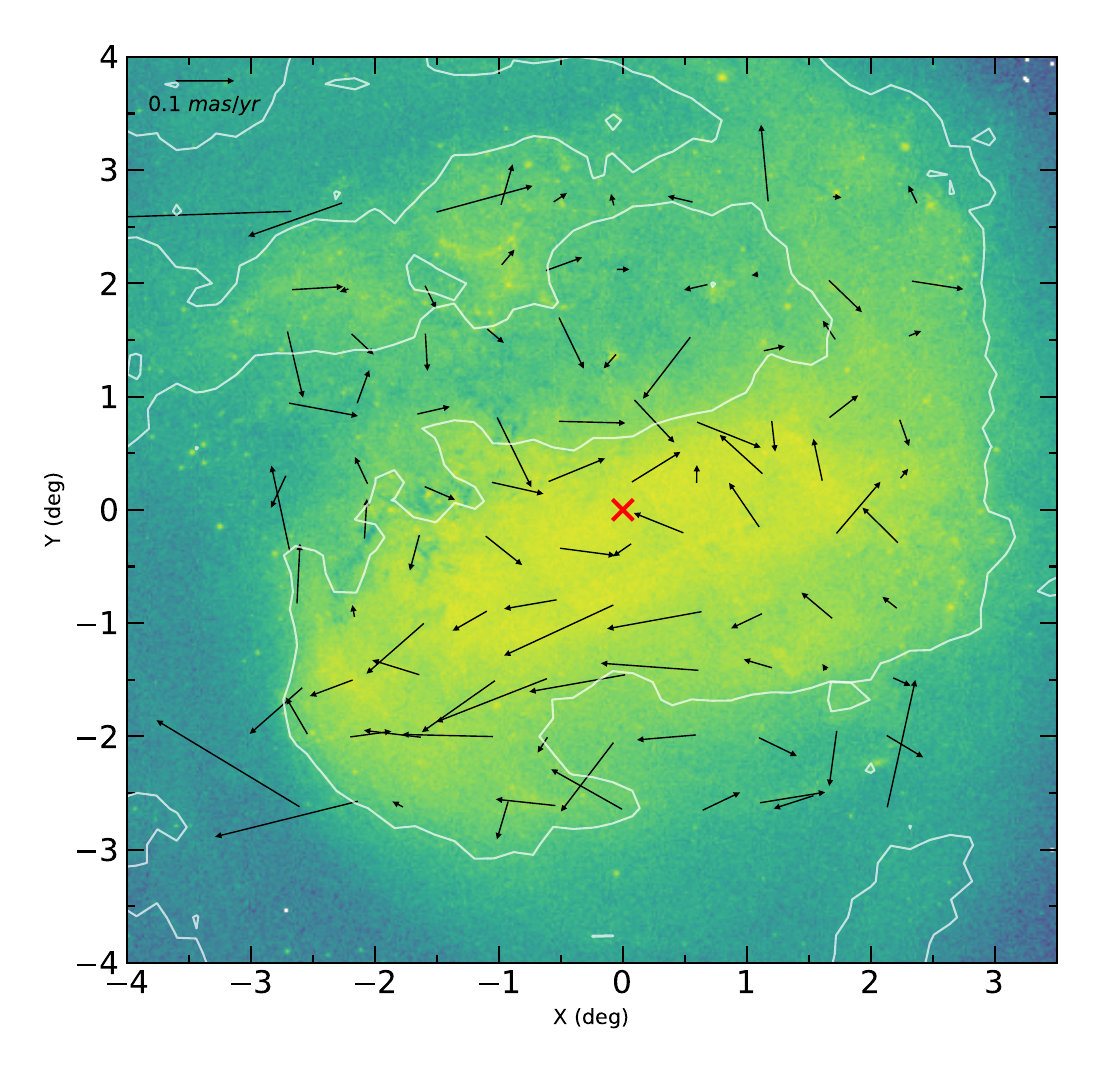}
    \caption{Same as in \ref{fig:vel_arrow_old}, but for the young population.}
    \label{fig:vel_arrow_young}
\end{figure*} 

\begin{figure*}[h!]
    \centering
    \includegraphics[width=9cm]{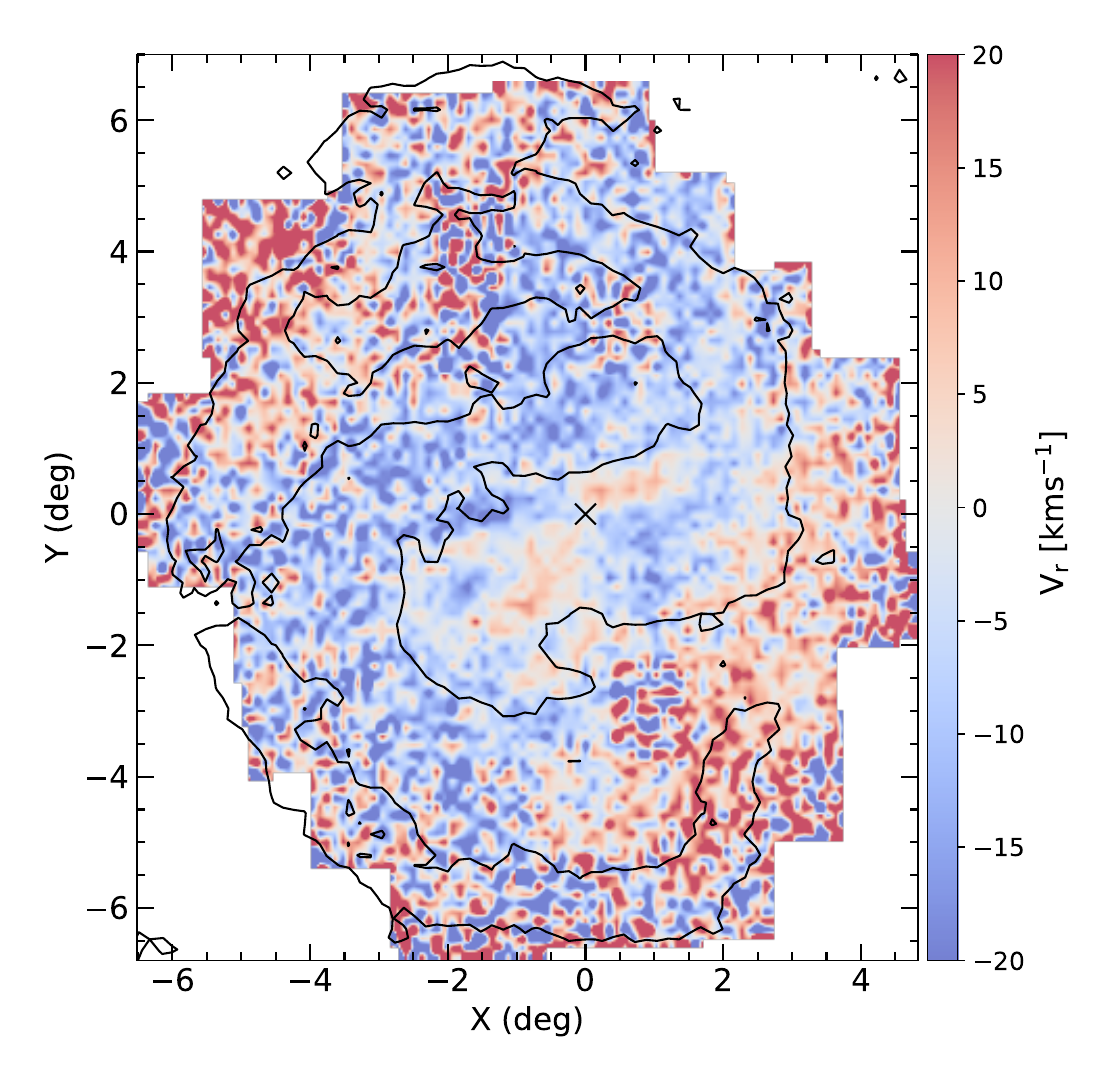}
    \hfill
    \includegraphics[width=9cm]{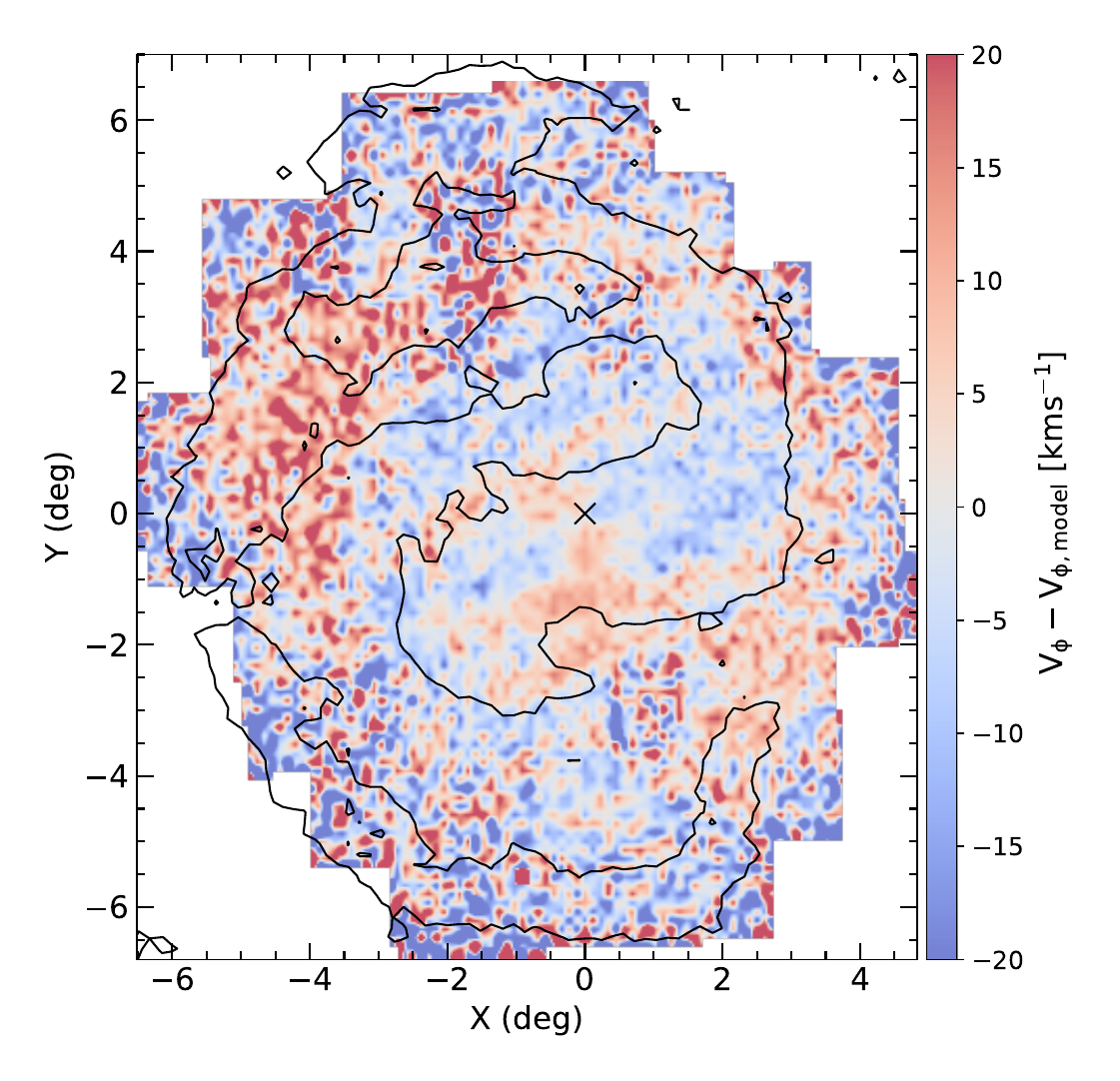}
    \caption{Same as in Figure \ref{fig:vel_polar}, but for the old population.}
    \label{fig:vel_polar_old}
\end{figure*} 

\begin{figure*}[h!]
    \centering
    \includegraphics[width=9cm]{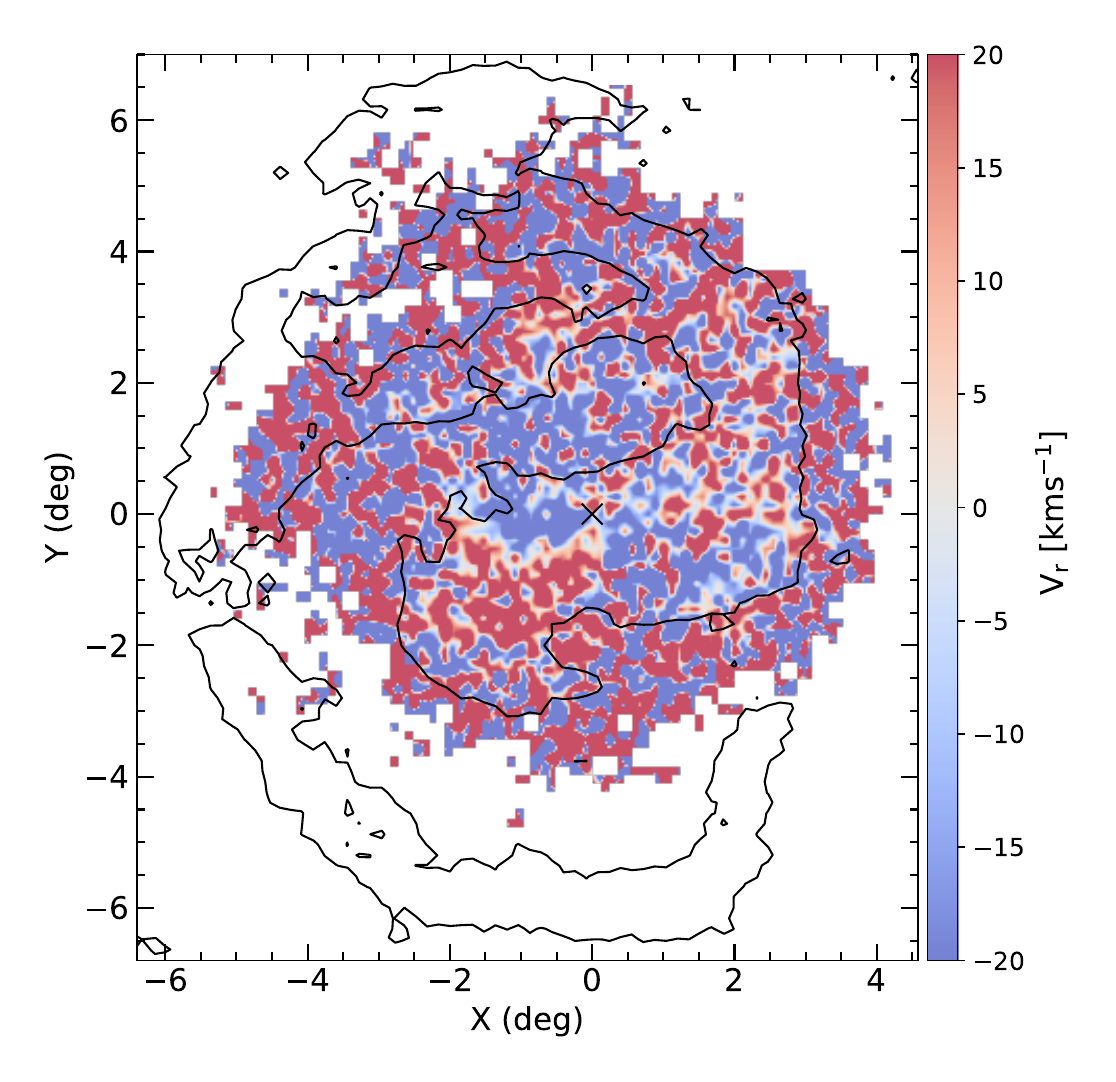}
    \hfill
    \includegraphics[width=9cm]{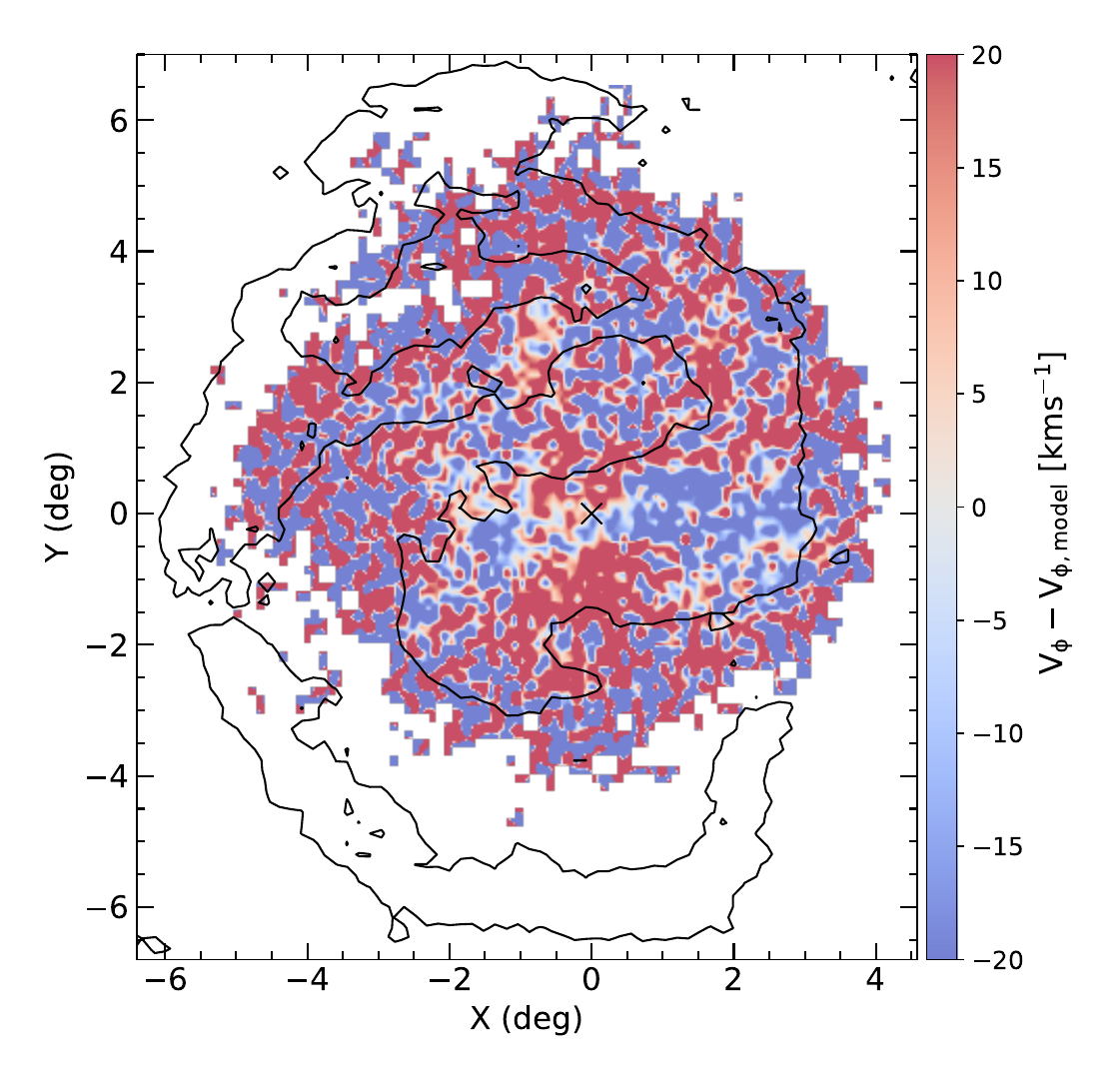}
    \caption{Same as in \ref{fig:vel_polar_old}, but for the young population.}
    \label{fig:vel_polar_young}
\end{figure*} 

\end{appendix}

\end{document}